\documentclass[%
reprint,
superscriptaddress,
nofootinbib,
nobibnotes,
 amsmath,amssymb,
 aps,
 prd,
 longbibliography,
 twocolumn,
]{revtex4-2}

\usepackage{float}
\usepackage[percent]{overpic}
\usepackage{graphicx, comment}
\usepackage[caption=false]{subfig}
\usepackage{dcolumn}
\usepackage{bm}
\usepackage{xcolor}
\usepackage[mathlines]{lineno}
\usepackage[normalem]{ulem}
\usepackage{amsfonts}
\usepackage{url}
\usepackage{lineno}
\usepackage{url}
\usepackage{float}
\usepackage{hyperref}
\usepackage[capitalise]{cleveref}

\begin{document}

\newlength{\figwidth}
\newlength{\fighalfwidth}
\setlength{\figwidth}{0.95\textwidth}
\setlength{\fighalfwidth}{0.5\textwidth}
\newcommand{\blue}[1]{{\textcolor{blue}{#1}}}
\newcommand{\red}[1]{{\textcolor{red}{#1}}}
\newcommand{\fixme}[1]{\textit{\textcolor{red}{Fixme: #1}}}

\newcommand{\Ar}{\mathrm{Ar}}
\newcommand{\nue}{\nu_e}
\newcommand{\antinue}{\bar{\nu}_e}


\title{Enhanced Reconstruction of Sub-GeV Neutrinos Charged Current Interactions in LArTPC}

\newcommand{\SBU}{Stony Brook University, SUNY, Stony Brook, NY 11794, USA}
\newcommand{\BNL}{Brookhaven National Laboratory, Upton, NY, 11973, USA}
\newcommand{\UTAustin}{University of Texas at Austin,
1 University Station, Austin, TX 78712, USA}

\author{Stone Chou} \affiliation{\SBU}
\author{Sanskar Jain} \affiliation{\UTAustin}
\author{Wei Shi} 
\email{wei.shi.1@stonybrook.edu} 
\affiliation{\SBU}
\author{Ciro Riccio} 
\email{ciro.riccio@stonybrook.edu}
\affiliation{\SBU}

\date{\today}

\begin{abstract}
This paper presents a comprehensive study of the reconstruction of sub-GeV neutrino charged-current interactions within a Liquid Argon Time Projection Chamber (LArTPC). We demonstrate that traditional charge-based calorimetry is fundamentally limited at sub-GeV scales by significant fluctuations in recombination and missing hadronic energy. We show that energy reconstruction using energy deposited as scintillation light ($L$) partially benefits from the self-compensating light effect. At neutrino energies above 400 MeV, the light-only reconstruction still outperforms charge-only methods that can separate EM and hadronic objects. The performance of the two remains comparable below 300 MeV. Using the energy-deposit information from both detector signals, we demonstrate at least 70\% efficiency can be achieved in tagging electron neutrinos and antineutrinos, respectively. By using a proximity-based algorithm coupled with a geometric lepton-exclusion cone, we also demonstrate the ability to isolate neutron-induced energy depositions from background. This enables an improvement of sub-GeV direction reconstruction by about 20 degrees for antineutrinos. This study provides new insights into how to enhance the physics reach of future LArTPC atmospheric neutrino analyses.
\end{abstract}

\maketitle


\section{Introduction}
\label{sec:intro}

Since the discovery of neutrino oscillation~\cite{SuperK_atm_nu_osc_discovery, SNO_discovery_1, SNO_discovery_2}, massive progress has been made in studying atmospheric neutrinos in current-generation experiments such as Super-Kamiokande~\cite{SuperK_2004result, SuperK_PhaseIresult, SuperK_PhaseI2IV_result, SuperK_PhaseI2V_purewaterresult}. Recent efforts culminate in their use to extract the neutrino mass ordering and the leptonic CP-violating (CPV) phase when jointly analyzed with beam neutrino data from T2K~\cite{T2K_SK_jointfit}. Sub-GeV atmospheric neutrinos are crucial for probing the CPV phase~\cite{subGeV_atmnuosc}. However, there has been difficulty reconstructing their direction, energy, and charge, which affects the calculation and simulation of the neutrino oscillation probability and weakens the physics sensitivity. 

Next-generation neutrino experiments, such as DUNE, will continue to use atmospheric neutrinos to probe the leptonic mixing sector. Once DUNE turns on its far detector at the end of this decade, each 17-kt module will collect hundreds of atmospheric neutrinos~\cite{DUNE_FD_TDR_Physics}. The LArTPC detector technology at DUNE offers new possibilities in the sub-GeV neutrino reconstruction. Previous studies have reported GeV and MeV neutrino reconstruction using charge and light signals in LAr~\cite{Selfcompensatinglight4GeV, MeV_nueCCLAr_light}, while the sub-GeV energy range study is missing. Meanwhile, DUNE recently reported the capabilities in reconstructing atmospheric neutrinos in its horizontal drift module~\cite{DUNE_HD_Atmnureco}. The report indicates that the resolution of both neutrino direction and energy reconstruction is significantly improved when incorporating more than just the lepton information. Most importantly, neutral particles such as neutrons constitute a significant part of the missing energy (or momentum) in neutrino interactions. Neutron identification and energy reconstruction via neutron-induced secondary particles are implemented in other detector technologies~\cite{superfgd}. It is timely to study more possibilities of sub-GeV neutrino reconstruction in an LArTPC. In this work, we explore ways to improve sub-GeV neutrino reconstruction by leveraging information that has not been extensively studied before, such as the light detection system and neutron energy deposition.

The paper is structured as follows: \Cref{sec:sim} describes the simulation sample used in this paper. The energy reconstruction performance using both the charge and light detector signals is discussed in \Cref{sec:energy}. The charge discrimination between neutrinos and antineutrinos is further discussed in \Cref{sec:separation}. \Cref{sec:dir} details the reconstruction of the direction of sub-GeV neutrinos using neutron energy deposits. The complementarity of this work to a recent work in Ref.~\cite{n_blip_reco_Wan} is discussed in \Cref{sec:discuss}. The final summary is presented in \Cref{sec:summary}.

\section{Simulation}
\label{sec:sim}

In this study, charged-current $\nu$-Ar interactions are simulated for each neutrino flavor ($\nu_e$, $\bar{\nu}_e$, $\nu_\mu$, $\bar{\nu}_\mu$), using the AR23\_20i tune~\cite{AR23tune} of the \texttt{GENIE} v3.04.00 neutrino event generator, in the sub-GeV range i.e. with incident neutrinos of energies ranging from 100 MeV (200 MeV) to 1000 MeV for $\nu_e$ and $\bar{\nu}_e$ ($\nu_\mu$ and $\bar{\nu}_\mu$), in 100 MeV increments. 1000 interactions were simulated per flavor per incident neutrino energy. The \texttt{edep-sim} software package, interfacing with \texttt{GEANT4} v10.6.p01 -- same as the setup used in Ref.~\cite{Selfcompensatinglight4GeV, MeV_nueCCLAr_light} -- is used to model the propagation and energy deposition of all subsequent primary and secondary particles in an infinite liquid argon volume.

\section{Sub-GeV Neutrino Energy Reconstruction}
\label{sec:energy}

Previous simulation studies have shown that light calorimetry in a LArTPC can play a unique role in neutrino energy reconstruction in multi-GeV~\cite{Selfcompensatinglight4GeV} and tens of MeV~\cite{MeV_nueCCLAr_light} energy ranges. This study seeks to fill the gap in the sub-GeV energy regime. The goal is to study the performance of various neutrino energy reconstruction methods similar to those presented in Ref.~\cite{Selfcompensatinglight4GeV} based on the simulated charge and light signals, with particular interest in assessing the benefits of utilizing the light signal, which will become more enhanced in upcoming LArTPC neutrino experiments such as DUNE~\cite{DUNE_Phase2}. 

We begin in \Cref{sec:Edep_process} with a discussion of energy deposition processes for sub-GeV neutrinos, highlighting the differences relative to the multi-GeV regime. We then review the energy reconstruction methods and show their performance in \Cref{sec:Ereco_method}. After these, we will interpret the results and comment on differences to the multi-GeV neutrino case in \Cref{sec:erecoperformance}.

\subsection{Sub-GeV neutrino energy deposition in LAr}
\label{sec:Edep_process}

We adopt the same definition as Ref.~\cite{Selfcompensatinglight4GeV} for the total available energy, $E_\text{avail}$, which includes the kinematic energy of protons and neutrons, and the total energy of electrons, pions, muons, and gammas. Smearing effects between $E_\nu$ and $E_\text{avail}$, which fluctuate from event to event, contribute to a spread in the $E_\text{avail}$ distribution. The AR23\_20i tune~\cite{AR23tune} used in this work models energy loss from the nucleon removal, which varies with the nucleon momentum. More significantly, some energy is gained or lost during final-state interactions (FSI). The hadrons produced in the initial neutrino interaction interact with other nucleons in the struck $^{40}$Ar nucleus, which could produce additional nucleons in the final state as well as occasional re-absorption of some primary hadrons. FSI is less pronounced in the multi-GeV neutrino interactions, but is crucial for sub-GeV neutrinos.

\begin{figure}[h!]
    \centering
    \includegraphics[width=1.0\linewidth]{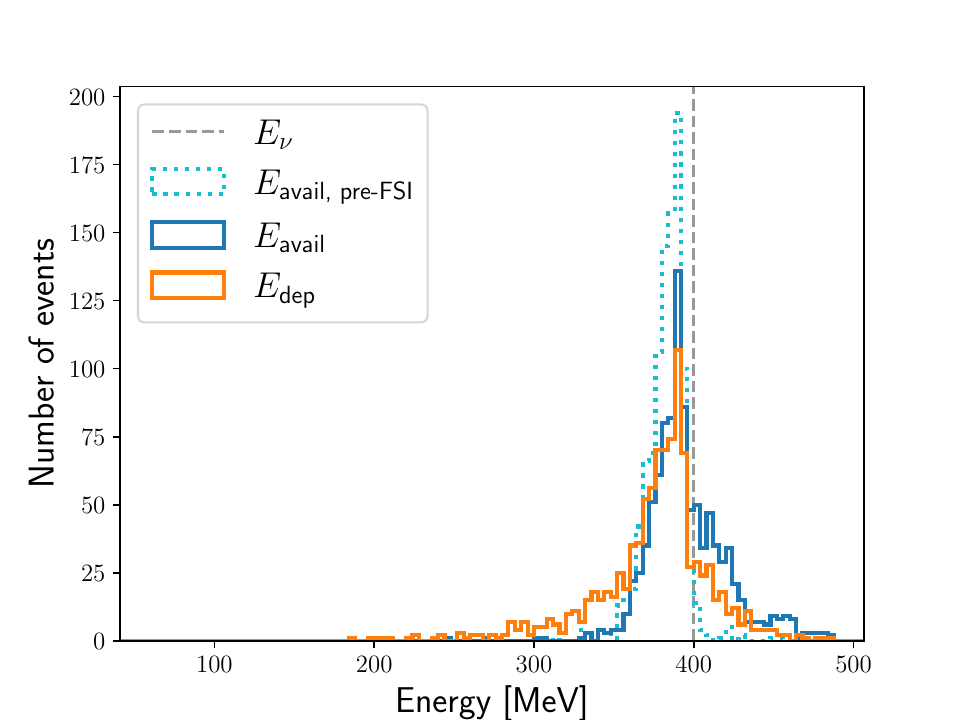}
    \caption{$E_\text{avail}$ (pre- and post-FSI) and $E_\text{dep}$ distributions for a sample of 1000 generated $\nu_e$ events with $E_\nu$ = 400 MeV. The bin width is 4 MeV.}
    \label{fig:Edep}
\end{figure}

The dotted light-blue histogram and the blue histogram in \cref{fig:Edep} show the distribution of the total available energy before and after the FSI, respectively, for a 0.4 GeV $\nu_e$ sample. As can be seen, many events (36\% of the sample) do not see any change in $E_\text{avail}$ before and after the FSI. These events also have the same list of particles at the primary vertex before and after the FSI, so it is safe to assume no FSI happens in these events. However, according to the model, for the majority of events, FSI increases $E_\text{avail}$ (52\% of events in the sample) rather than decreasing it (12\%). While the fraction of events that see no FSI appears to remain approximately 30\% across the sub-GeV $E_\nu$ range, the fraction of events where FSI raises $E_\text{avail}$ decreases from $\sim$70\% at $E_\nu=100$ MeV to $\sim$50\% at $E_\nu=1000$ MeV. Since the exact relationship between the simulated $E_\text{avail}$ and the true $E_\nu$ is highly model-dependent, we will focus on estimating $E_\text{avail}$ from the simulated detector signals, which is a reasonable approximation for the true $E_\nu$. 

\begin{figure} [h!]
    \centering
    \includegraphics[width=\linewidth]{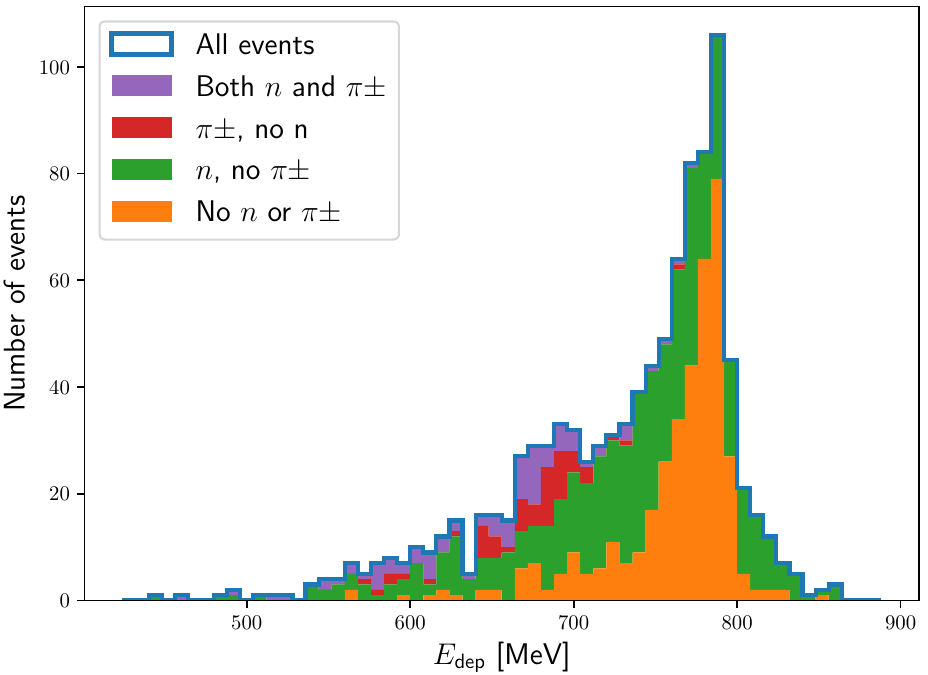}
    \caption{The distribution of deposited energy for the 1000 $E_\nu=800$ MeV events, broken down into different subsets of events depending on the presence or absence of primary neutrons or charged pions in their final states. The bin width is 8 MeV. Note that the majority of the tail to the left (i.e., to lower deposited energies) consists of events with (primary) neutrons or charged pions.}
    \label{fig:missing energy}
\end{figure}

After FSI, all outgoing particles deposit their energy in LAr by either ionization or excitation interactions with $^{40}$Ar atoms. However, not all $E_\text{avail}$ is deposited. The orange histogram in \cref{fig:Edep} shows the distribution of the total deposited energy, $E_\text{dep}$, for the 0.4 GeV sample. The $E_\text{dep}$ distribution has a much longer tail toward lower energies due to varying amounts of missing energy. The most prominent sources of missing energy are from neutrons and charged pions. Neutrons, being electrically neutral, cannot ionize or excite nearby $^{40}$Ar atoms to deposit their energy. Instead, depending on their kinetic energy, they interact with $^{40}$Ar nuclei through elastic and inelastic scattering, transferring part of their energy to recoiling nuclei or producing secondary particles, such as de-excitation $\gamma$ rays and additional nucleons. Inelastic scattering also results in some of the neutron's kinetic energy being lost to the binding energies required to break up the nucleus and release secondary hadronic particles. The amount of energy lost in nuclear breakup can vary significantly from event to event. At higher energies, neutrons can also induce spallation and other nuclear reactions. While charged pions can continuously deposit energy in LAr, they decay with some $E_\text{avail}$ carried away by neutrinos. \Cref{fig:missing energy} shows that events with neutrons and pions are mostly responsible for the tail in the $E_\text{dep}$ distribution we attributed to missing energy. 

Particles deposit energy in the LAr by ionizing or exciting $^{40}$Ar atoms. In this simplified model, a constant fraction $\beta=0.83$ of the deposited energy is assumed to produce ionization electrons, while the remaining fraction is converted into scintillation photons. The recombination of ionization electrons and argon ions results in some electrons essentially being exchanged for scintillation photons. As such, the actual fraction of quanta that end up as ionization electrons is $\beta R_c$, where $R_c$ is the charge recombination factor and depends on the electric field of the LArTPC and the $dE/dx$ of the specific particle. $R_c$ is calculated in the simulation using the Birks' Model as implemented in Ref.~\cite{Selfcompensatinglight4GeV}. Taking this into account, one can break up the total deposited energy $E_\text{dep}$ as a sum of two components: the energy in ionization electrons, i.e., charge,
\begin{equation}
Q\equiv \beta R_c E_\text{dep},    
\end{equation}
and the energy in scintillation photons, i.e., light, 
\begin{equation}
L\equiv (1-\beta R_c) E_\text{dep}.    
\end{equation}

\begin{figure}[h!]
    \centering
    \includegraphics[width=1.0\linewidth]{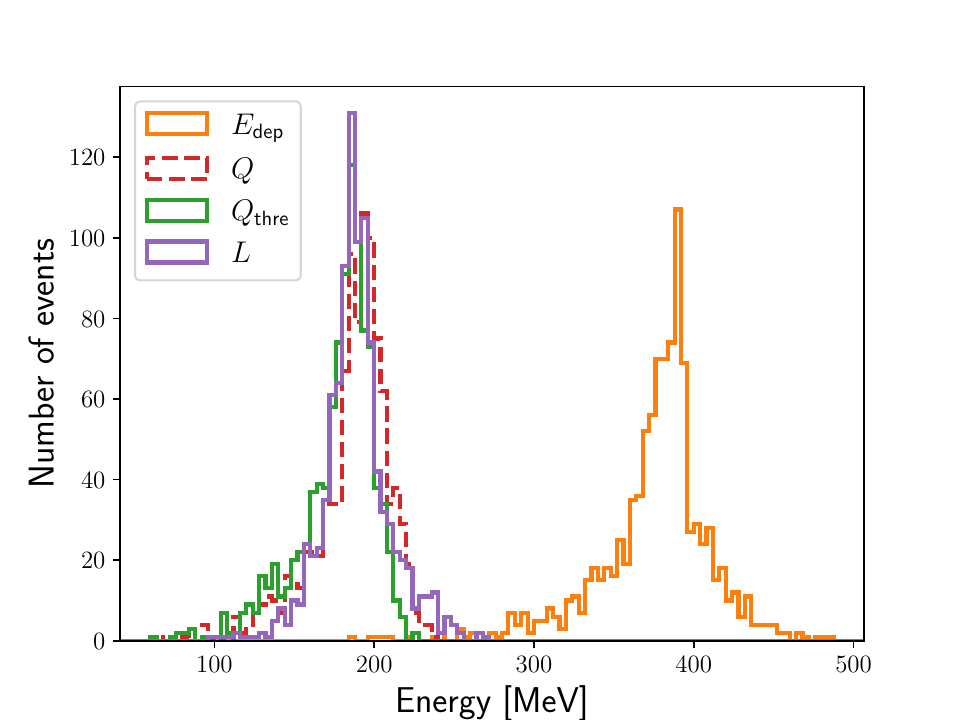}
    \caption{$E_\text{dep}$, $Q$ and $L$ distributions for a sample of 1000 generated $\nu_e$ events with $E_\nu = 400$ MeV. $Q_{thre}$ (also called $Q_{75}$) is the $Q$ obtained when a 75 keV threshold is applied to each ionization-charge deposition. The bin width is 4 MeV.}
    \label{fig:Edep2}
\end{figure}

Charge ($Q$) and light ($L$) are the detector signals that are used to reconstruct the incident neutrino energy. The $Q$ and $L$ distributions are shown in the red and purple histograms, respectively, in \cref{fig:Edep2}. Here, we also consider an optimistic 75 keV charge detection threshold to reflect that detector readout noise sets a minimum limit on the amount of charge that can be reconstructed. The $Q$ distribution after applying this threshold ($Q_{\text{thre}}$ or interchangeably $Q_{75}$) is shown in the green histogram of \cref{fig:Edep2}. While the distribution of $Q$ maintains roughly the same shape as $E_\text{dep}$, its peak is less sharp due to the recombination effect. For the light collection, we have assumed a uniform photon-collection efficiency of 0.83\%, corresponding to an average light yield of 180 photoelectrons per MeV for a MIP. The expected number of detected photons and the final detected energy deposited in light follow the same treatment presented in Ref.~\cite{MeV_nueCCLAr_light, Selfcompensatinglight4GeV}. The $L$ distribution is less smeared out and more symmetric than the $E_\text{dep}$ distribution. The reasons for this will be addressed in \cref{sec:Ereco_method} below. 

\begin{figure} [h!]
    \centering
    \includegraphics[width=\linewidth]{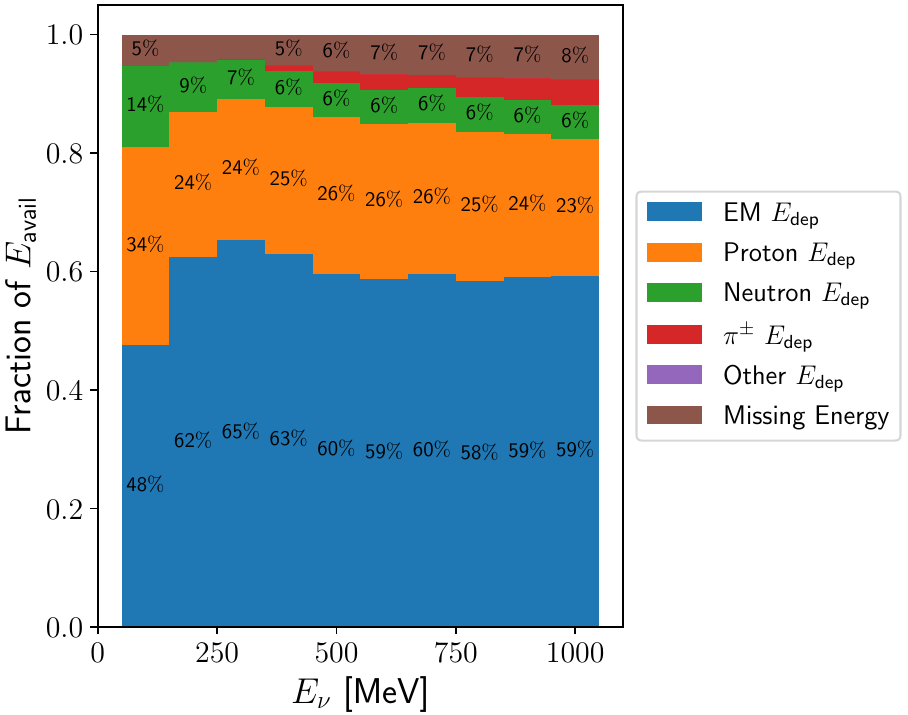}
    \caption{Fraction of total available energy $E_\text{avail}$ distributed among final state particles on average in simulated sub-GeV $\nu_e$ samples.}
    \label{fig:edep_breakdown}
\end{figure}

At this point, it is important to summarize the differences between the energy share profiles of sub-GeV and multi-GeV $\nu_e$ CC events. In Ref.~\cite{Selfcompensatinglight4GeV}, it was found that only 10.4\% of the $E_\text{avail}$ in 3 GeV $\nu_e$ events is deposited by protons and their descendants. By contrast, this fraction increases to 24.9\% in sub-GeV events (\cref{fig:edep_breakdown}). This fraction remains stable in the 23\%-26\% range across all 200-1000 MeV $\nu_e$ samples. This increase is compensated for by having a smaller fraction of $E_\text{avail}$ deposited in pions (2.7\% versus 9.8\%). Below $E_\nu=400$ MeV, essentially no pions are created. There is a slight decrease in the fraction deposited by the EM component (59.6\% versus 64.5\%). The fraction of missing energy is also slightly less in the sub-GeV regime (6.6\% on average instead of 8.0\%). Also, essentially no other heavier particles (which made up 1.3\% of the 3 GeV $E_\text{avail}$) are created.     

In addition to contributing a larger share of the deposited energy, the final-state protons in the sub-GeV regime also have a smaller average percentage of missing energy than the 3 GeV events (4.5\% versus 11.8\%). Primary neutrons behave similarly in both energy regimes, with 38\% missing energy for sub-GeV events and 36\% for 3 GeV. The fraction of missing energy increased for primary charged pions (34.5\% versus 24.7\%). This is expected since pions are created with smaller $E_\text{avail}$ in sub-GeV neutrino events. However, primary pions are only in $\approx$$8\%$ of sub-GeV events. The percentage of missing energy in the EM component is consistently negligible across energy regimes (0.04\%).    

Overall, the difference setting sub-GeV events apart from the multi-GeV case is the larger role of the primary protons in the deposited energy: a greater fraction of the events' $E_\text{avail}$ is apportioned to protons, and most of the proton kinetic energy is deposited. This has significant implications for the quality of energy reconstruction achievable in this regime, because the charge recombination factor $R_c$ for low-energy protons is much lower than that for high-energy MIP protons and depends on their $dE/dx$.

\subsection{Methods of energy reconstruction}
\label{sec:Ereco_method}

\begin{figure}[h]
    \centering
    \includegraphics[width=1\linewidth]{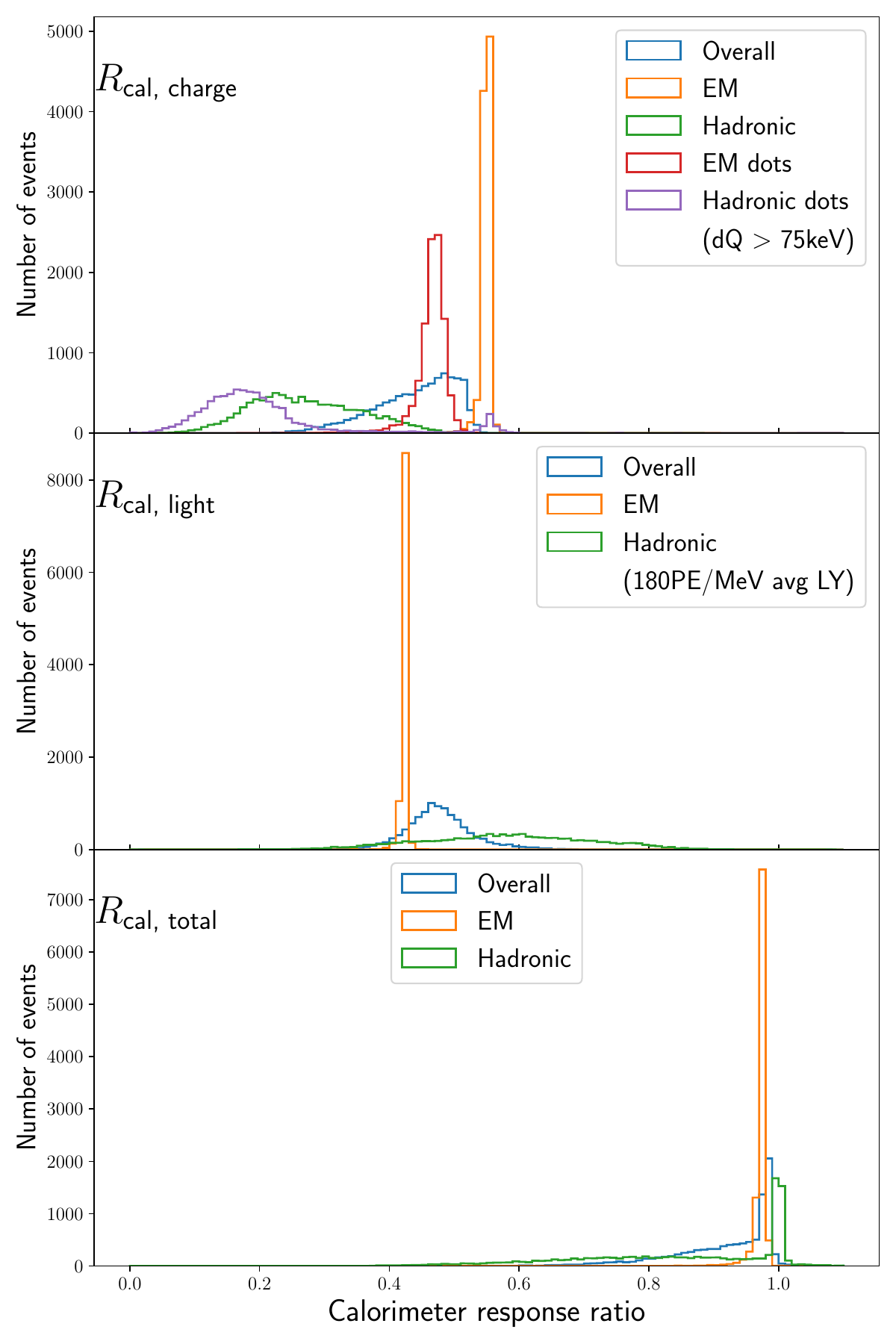}
    \caption{Calorimetric response distributions for the sample of $10^4$ generated $\nu_e$ events. $Q$ values are obtained assuming a charge threshold of 75 keV ($Q_{75}$), and $L$ values are modeled using an average light yield (LY) of 180 photoelectrons per MeV of deposited energy by a MIP. All histograms have a bin width of 0.01. Note that the distributions for $R_\text{cal, charge}^{e(h) \text{ dots}}$ only include events with a non-zero amount of dot-like charge depositions.}   
    \label{fig:rcal}
\end{figure}

As discussed above, only a fraction of a particle's $E_\text{avail}$ is actually visible to a calorimeter in the form of $Q$ or $L$. This fraction is defined as the calorimetric response as in Ref.~\cite{Selfcompensatinglight4GeV},
\begin{equation}
R_\text{cal}\equiv\frac{E_\text{vis}}{E_\text{avail}},    
\end{equation}
where $E_\text{vis}$ is the energy deposited as charge ($Q$) or light ($L$) signals in the simulation. Naturally, if we knew $R_\text{cal}$ for a neutrino event, we could divide out the recorded visible energy by $R_\text{cal}$ to get the true available energy. However, the calorimetric response from individual outgoing particles can fluctuate significantly. The charge and light calorimetric responses are plotted in \cref{fig:rcal} for different groups of particles in sub-GeV neutrino events. We have previously noted that sub-GeV events have a larger fraction of their deposited energy in protons, meaning that lower-energy protons constitute a larger fraction of the hadronic component in the sub-GeV regime than in the multi-GeV case. We also note that the charge recombination factor $R_c$ for protons decreases steeply with the initial proton kinetic energy below $\approx$200 MeV (cf. \cref{fig:charge-responses,fig:light-responses}). Thus, on average, the charge calorimetric response is lower and the light calorimetric response correspondingly higher for the hadronic component of sub-GeV $\nu_e$ events than for that of multi-GeV events. By contrast, the EM-component calorimetric responses for sub-GeV and multi-GeV $\nu_e$ events are essentially identical. 

From \cref{fig:rcal}, we extract the peak calorimetric response in each distribution as scaling factors that would, when divided by, approximately scale the peak in the visible energy signal -- whether that be $Q$, $L$, or $Q+L$ -- to the peak in the distribution of $E_\text{avail}$ for a sample of events. Below, we consider five major methods of energy reconstruction. Most of them follow Ref.~\cite{Selfcompensatinglight4GeV}, except we consider one more method of combining charge and light signals (Q+L):

\begin{itemize}
    \item L1: a neutrino energy reconstruction based solely on light signals. We divide out the energy deposited as light $L$ by the peak overall light calorimetric response $R_\text{cal, light}$ found from the simulation sample,
    \begin{equation}
    E_\text{rec}=L/0.465,   
    \end{equation}

    \item Q1: a neutrino energy reconstruction based solely on charge signals. Similarly, we divide out the energy deposited as charge $Q_{75}$ by the peak overall charge calorimetric response $R_\text{cal, charge}$,
    \begin{equation}
    E_\text{rec}=Q_{75}/0.485,
    \end{equation}

    \item Q+L: a neutrino energy reconstruction based on the summed energy deposited in charge and light. We divide out the sum of the energies deposited as charge and light signals $Q_{75}+L$ by the peak overall combined calorimetric response $R_\text{cal, total}$, 
    \begin{equation}
    E_\text{rec}=(Q_{75}+L)/0.984,
    \end{equation}

    \item Q2: a neutrino energy reconstruction based solely on charge signals with the assumption that the charge signals from the hadronic and EM components can be separated in the detector. We will divide out the energy in charge depositions originating from EM primary particles only -- $Q_{75,e}$ -- by the peak EM-component charge calorimetric response $R_\text{cal, charge}^{e}$ to reconstruct the EM-component available energy, and add it to the energy in charge depositions originating from hadronic particles only -- $Q_{75,h}$ -- divided by the peak hadronic-component charge calorimetric response $R_\text{cal, charge}^{h}$,
    \begin{equation}
    E_\text{rec}=\frac{Q_{75,e}}{0.553} + \frac{Q_{75,h}}{0.225},
    \end{equation}

    \item Q3: a neutrino energy reconstruction based solely on charge signals with the assumption that the track objects can be separated from blip-like objects, and we can also differentiate blip-like objects between those from EM primary particles and hadronic primary particles. This is the ideal case, but it is challenging to realize in experiments today. We assert that energy for all particles whose tracks are longer than 2 cm can be faithfully reconstructed. Then, the available energy can be reconstructed by dividing out the energies in the EM- and hadronic-component dot-like charge depositions $Q_{75,e}^\text{dots}$ and $Q_{75,h}^\text{dots}$ respectively by the respective peak calorimetric responses $R_\text{cal, charge}^{e(h) \text{ dots}}$ and adding them together with the total measured energy deposited by tracks:
    \begin{equation}
    E_\text{rec}=E_\text{dep}^\text{tracks} + \frac{Q_{75,e}^\text{dots}}{0.475} + \frac{Q_{75,h}^\text{dots}}{0.165}
    \end{equation}

\end{itemize}

\begin{figure}[h]
    \centering
    \includegraphics[width=1\linewidth]{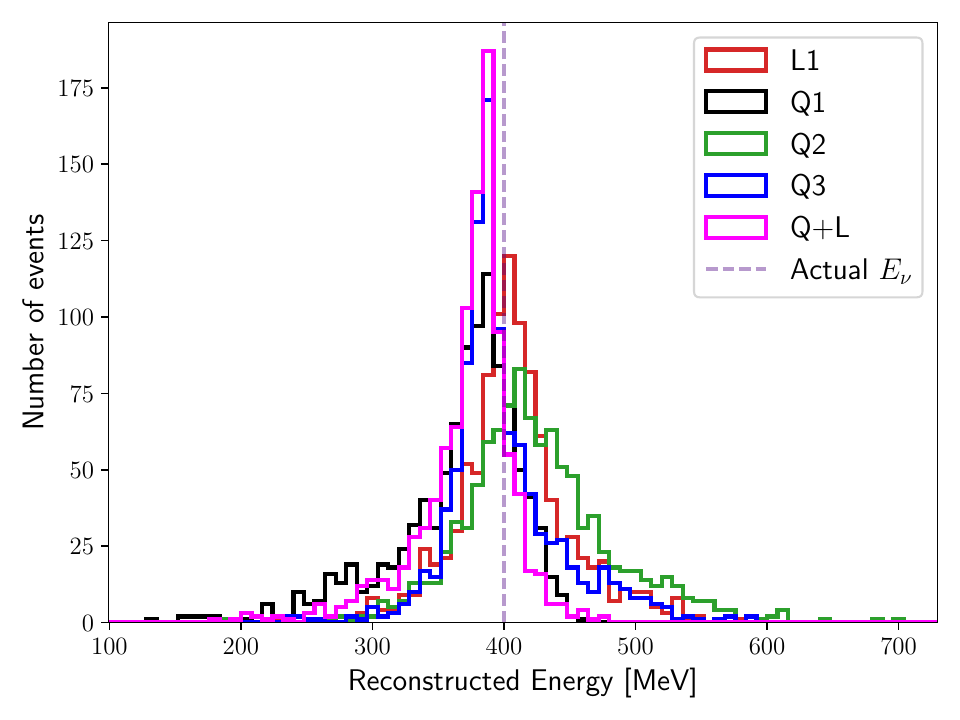}
    \caption{Distributions of the reconstructed incident neutrino energies $E_\text{rec}$ obtained for the sample of 1000 $\nu_e$ events with true $E_\nu=400$ MeV using the five different methods detailed above. The bin width is 8 MeV.}
    \label{fig:recoEdistribution}
\end{figure}

\begin{figure}[h]
    \centering
    \includegraphics[width=1\linewidth]{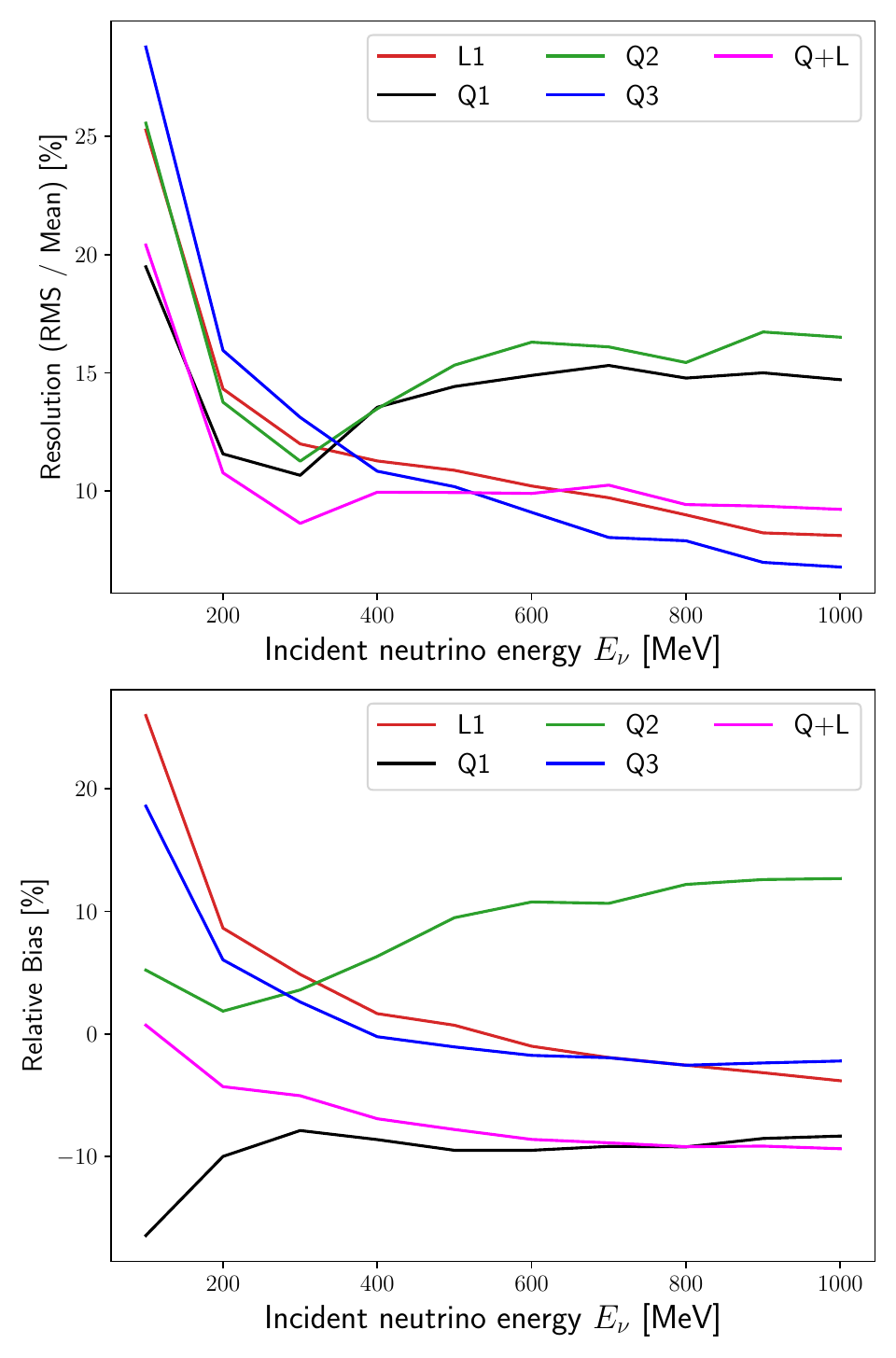}
    \caption{Energy resolution $\sigma_\text{(RMS)}/\overline{E_\text{rec}}$ (top) and relative energy biases $(\overline{E_\text{rec}}-E_\nu)/E_\nu$ (bottom) achieved by the different energy reconstructions for $\nu_e$ events in each of the ten $10^3$-event samples, which span $E_\nu=100$ MeV to $E_\nu=1000$ MeV. The closer a resolution or bias value is to zero, the better the performance it represents.}
    \label{fig:resolution}
\end{figure}

We apply all methods to each sub-GeV neutrino sample and obtain a distribution of the reconstructed available energy $E_\text{rec}$ from each method. One example is shown in \cref{fig:recoEdistribution} for the 0.4 GeV $\nu_e$ sample. We then calculate the resolution and relative bias achieved by each method, where resolution is defined as the RMS of the $E_\text{rec}$ distribution divided by the mean reconstructed energy $\overline{E}_\text{rec}$ and the relative bias is defined as $\frac{\overline{E}_\text{rec}-E_\nu}{E_\nu}$ i.e. the relative difference between the mean reconstructed energy and the true incident $\nu_e$ energy. The resolution and relative biases of the five methods are plotted as a function of $E_\nu$ in \cref{fig:resolution}.

\subsection{Discussion of sub-GeV energy reconstruction performance}
\label{sec:erecoperformance}

\begin{figure}
    \centering
    \includegraphics[width=\linewidth]{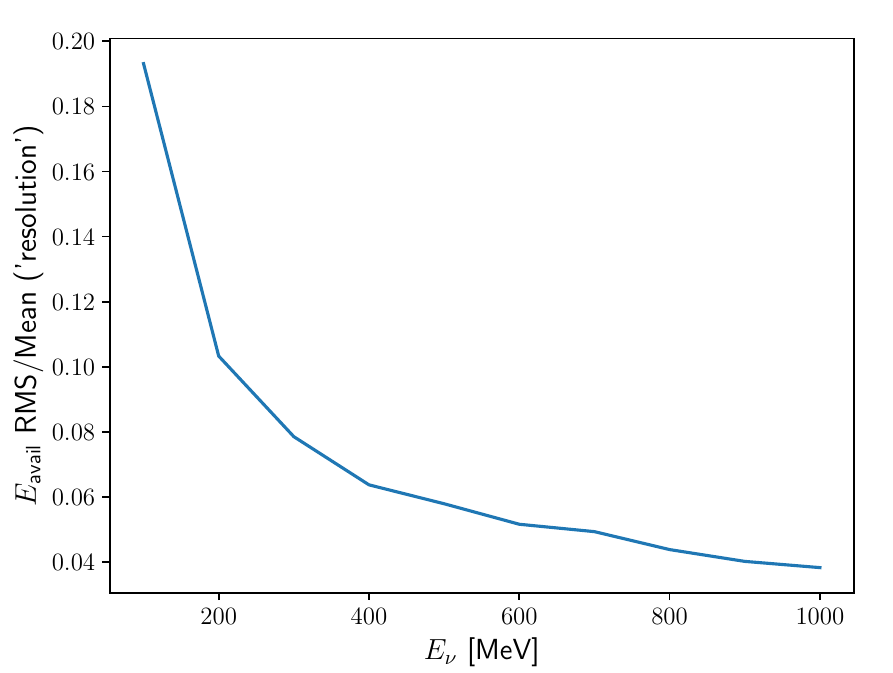}
    \caption{Resolutions of the simulated true $E_\text{avail}$, which represent the best achievable performance of all energy reconstruction methods.}
    \label{fig:eavail_resolution}
\end{figure}

All five methods ultimately attempt to reconstruct the total available energy of the event. It stands to reason that the best achievable performance of the energy reconstruction methods described above would be the resolutions of the simulated true $E_\text{avail}$ itself. This is shown in \cref{fig:eavail_resolution}. It shows that the resolution of the true available energy improves monotonically with $E_\nu$. The L1 and Q3 reconstruction results in \cref{fig:resolution} nicely preserve this trend. However, at higher $E_\nu$, the resolutions of Q1 and Q2 diverge significantly from this trend. Below, we discuss the performance of each of the five reconstruction methods.

\subsubsection*{Q1}

\begin{figure}
    \centering
    \includegraphics[width=\linewidth]{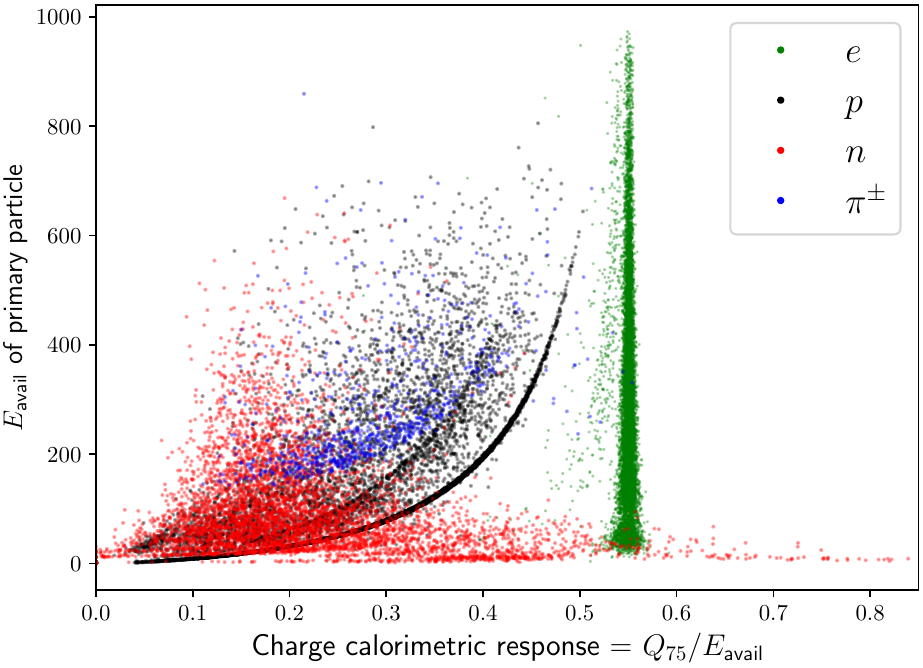}
    \caption{Scatter plot of the charge calorimetric responses of different primary particles against their available energy across all $10^4$ simulated events (with $E_\nu$ ranging from 100 MeV to 1000 MeV). The responses are calculated accounting for all descendants of the primary particles. Notice that all hadronic particles have significantly smaller charge-calorimetric responses than the primary electrons (EM component).}
    \label{fig:charge-responses}
\end{figure}

To understand why Q1 has degraded resolution at higher $E_\nu$ in \cref{fig:resolution}, we can look at the charge calorimetric response $R_\text{cal, charge}$ for different kinds of primary particles as a function of their available energy, as shown in \cref{fig:charge-responses}. We observe first that almost all possible $R_\text{cal, charge}$ values of hadronic particles ($p, n, \pi^\pm$) are well below the possible $R_\text{cal, charge}$ values of the electron (which are tightly constrained around 0.55). This is because of the high recombination effect for low-energy protons due to their large dE/dx (suppressing their $Q$ in favor of $L$), and larger amounts of missing energy for neutrons and pions (diminishing both their $Q$ and $L$). Because of this, the true \textit{overall} $R_\text{cal, charge}$ will fluctuate significantly from event to event with different splits of $E_\text{avail}$ between the components. This limits the resolution achievable when we uniformly divide all events' total $Q$ by just the peak of the $R_\text{cal, charge}$ distribution, as done in Q1.

As we go to higher $E_\nu$, a wider range of available energies becomes possible for the primary particles. Note that, as can be seen in \cref{fig:charge-responses}, at lower energies, the $R_\text{cal, charge}$ values for protons and neutrons tend to be in a similar low range. Therefore, whether or not an event even has a primary neutron and how the $E_\text{avail}$ is distributed between protons and neutrons makes little difference to the overall $R_\text{cal, charge}$. However, as we move to higher $E_\nu$, higher-energy primary protons and neutrons are produced. In this case, the distributions begin to diverge, with primary neutrons on average having significantly smaller $R_\text{cal, charge}$ than primary protons. This causes the event-to-event fluctuation in the overall $R_\text{cal, charge}$ to increase with rising $E_\nu$, to the extent that the Q1 reconstruction fails to reconstruct $E_\text{avail}$ well for higher $E_\nu$ in \cref{fig:resolution}.

\subsubsection*{L1}

\begin{figure}
    \centering
    \includegraphics[width=\linewidth]{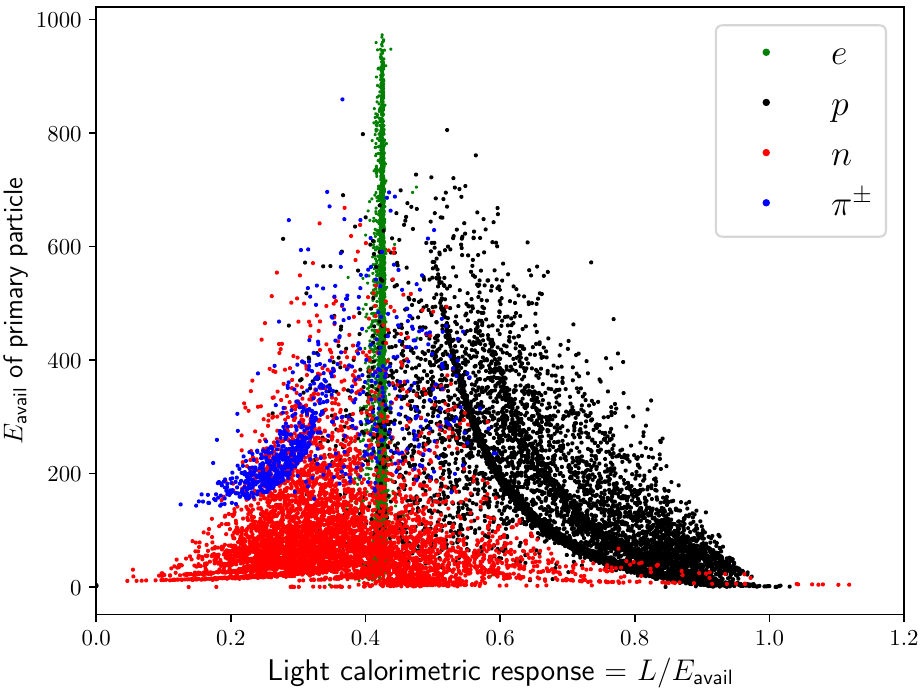}
    \caption{Scatter plot of the light calorimetric responses of different primary particles against their available energy across all $10^4$ simulated events ($E_\nu$ ranging from 100 MeV to 1000 MeV). The responses are calculated accounting for all descendants of the primary particles. Notice how the hadronic particles' light calorimetric responses are centered close to that of the primary electrons (EM component), and that at higher available energies, the light responses of the different hadronic particles all actually converge towards that of the EM component, similar to the self-compensating light observed in Ref.~\cite{Selfcompensatinglight4GeV}.}
    \label{fig:light-responses}
\end{figure}

The above discussion on Q1 begs the question of why the L1 method does appear to faithfully reconstruct the $E_\text{avail}$ at higher $E_\nu$, given its resolution continues to monotonically decrease with $E_\nu$. Again, we can understand the behavior of this method better using the scatter plot of the light calorimetric response $R_\text{cal, light}$ values for different primary particles in \cref{fig:light-responses}. Here, we see that unlike the charge calorimetric responses, the average electron $R_\text{cal, light}$ and the average hadronic particle's $R_\text{cal, light}$ are much closer to each other -- even for these sub-GeV events, there is to some extent \textit{self-compensation in light calorimetry}~\cite{Selfcompensatinglight4GeV} between the calorimetric responses of the EM and hadronic components. This is because, unlike in the case of charge, the stronger recombination for primary protons \textit{increases} their $R_\text{cal, light}$ above that of electrons, while the greater missing energy for neutrons and pions still lowers their $R_\text{cal, light}$ below that of electrons. Consequently, the fluctuation in the overall $R_\text{cal, light}$ due to varying partitions of the total available energy into the EM and hadronic components is greatly reduced, compared to charge. 

This is consistent with how, in \cref{fig:rcal}, the overall $R_\text{cal, light}$ peak is close to both the EM- and hadronic-component peaks, while the overall $R_\text{cal, charge}$ peak is far from the hadronic-component peak. The self-compensation in $L$ does not reach the extent seen in multi-GeV events, where the EM- and hadronic-$R_\text{cal, light}$ essentially peak at the same value. It is nevertheless sufficient in the sub-GeV case that, even using only the light signal, the relatively simple L1 method achieves performance comparable to that of the more complex Q3 method based on charge signals.

Additionally, at higher energies, the $R_\text{cal, light}$ values of the different hadronic particles converge toward that of the electron, meaning the light self-compensation effect strengthens with increasing energy (consistent with its prevalence in the multi-GeV case~\cite{Selfcompensatinglight4GeV}). Consequently, the presence of varying hadronic primary particles, and the energy partition among them, causes less event-to-event fluctuation in the overall $R_\text{cal, light}$ at higher energies. It therefore makes sense that L1 continues to reconstruct $E_\text{avail}$ well and thus continues to improve in resolution at higher $E_\nu$ in \cref{fig:resolution}.

The worse resolution of L1 compared to Q1 at the lowest $E_\nu$ is also expected: the $R_\text{cal, light}$ values of the lowest-energy primary hadronic particles span the entire range from 0 to 1, whereas the corresponding $R_\text{cal, charge}$ values do not.

\subsubsection*{Q+L}

Ref.~\cite{Selfcompensatinglight4GeV} also suggested that Q+L does not offer any significant advantage over Q1 for multi-GeV neutrino events because the Q+L reconstructed energy distribution inherits the missing energy long tail from Q1. However, as seen in \cref{fig:resolution}, Q+L performs well across the sub-GeV range, achieving comparable resolutions to L1 and Q3 even at the higher $E_\nu$. Given that, in the sub-GeV events, the smearing due to recombination is a much more pronounced source of variance in Q1, adding $Q_{75}$ and $L$ together to eliminate this smearing naturally yields a greater benefit. In addition, we found in \Cref{sec:Edep_process} that sub-GeV events overall have less missing energy on average: more of the available energy gets allocated to protons -- which tend to deposit close to 100\% of their energy -- as opposed to pions (a large source of missing energy). It is therefore reasonable that Q+L performs better in the sub-GeV than the multi-GeV regime.

\subsubsection*{Q2 and Q3}

\begin{figure}
    \centering
    \includegraphics[width=0.9\linewidth]{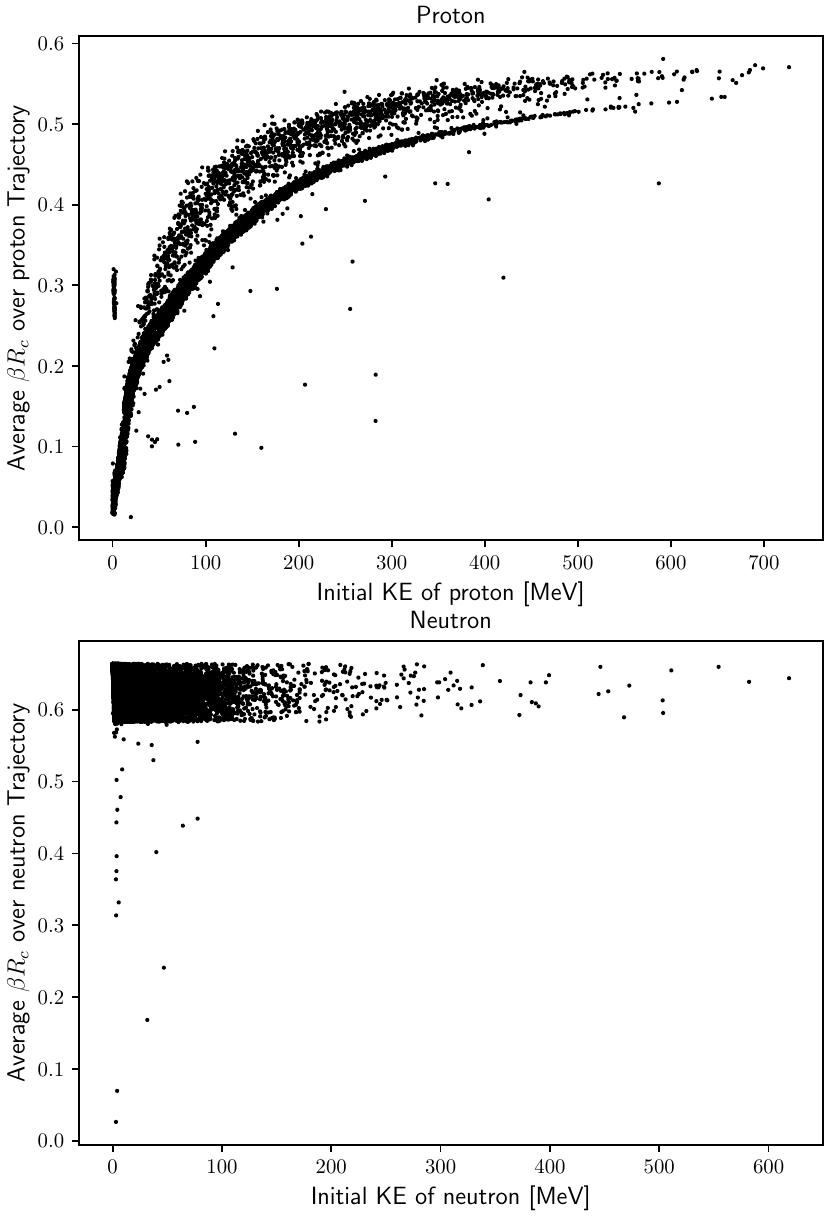}
    \caption{While the average charge recombination factor $R_c$ of proton tracks (top) varies steeply with proton energy (particularly at low energies), the $R_c$ of neutrons (bottom) is mostly constrained to a constant, strict range for all energies. The protons' wide range of possible $R_c$ values results in a large spread in hadronic $R_\text{cal, charge}$ and, consequently, degraded Q2 performance at higher $E_\nu$. This can be compensated for if each proton track's $R_c$ value could be determined, as in Q3. Note that the energy depositions which occur when a proton inelastically scatters on an $^{40}$Ar at the end of its track must have a large constant $R_c$, which artificially raises the average $R_c$ for those (less-common) proton trajectories, forming a second parallel "band" above the main curve. This plot includes every proton and neutron trajectory across all $10^4$ simulated $\nu_e$ events.}
    \label{fig:proton dE/dx}
\end{figure}

The crux of the reason why Q2 fails to reconstruct $E_\text{avail}$ well at high $E_\nu$ (and thus has a worsening resolution beyond $E_\nu=300$ MeV) but Q3 does not is the fact that proton tracks in sub-GeV $\nu_e$ events have a large variation in their $R_\text{cal, charge}$ values. Despite the fact that almost all proton tracks consistently deposit essentially 100\% of their available energy into the LAr, proton tracks of different initial kinetic energies have significantly different fractions of their deposited energy in charge ($Q$) as opposed to light ($L$) due to differing amounts of recombination (\cref{fig:proton dE/dx,fig:charge-responses}). The average $R_c$ of a proton track increases from close to 0 (all deposit energy goes into light) at kinetic energies near 0 MeV to approaching around 0.55 at the highest achieved kinetic energies, with the relationship becoming increasingly steep at lower kinetic energies.

At higher $E_\nu$, each primary proton created in an event has a wider range of permissible available energies to start with -- from 0 to a little less than $E_\nu$ -- and so each primary proton has a wider range of possible $R_\text{cal, charge}$ values. This makes scaling the hadronic-component $Q_{75}$ values obtained across many events by any single factor increasingly ineffective as $E_\nu$ increases.  

\begin{figure}
    \centering
    \includegraphics[width=\linewidth]{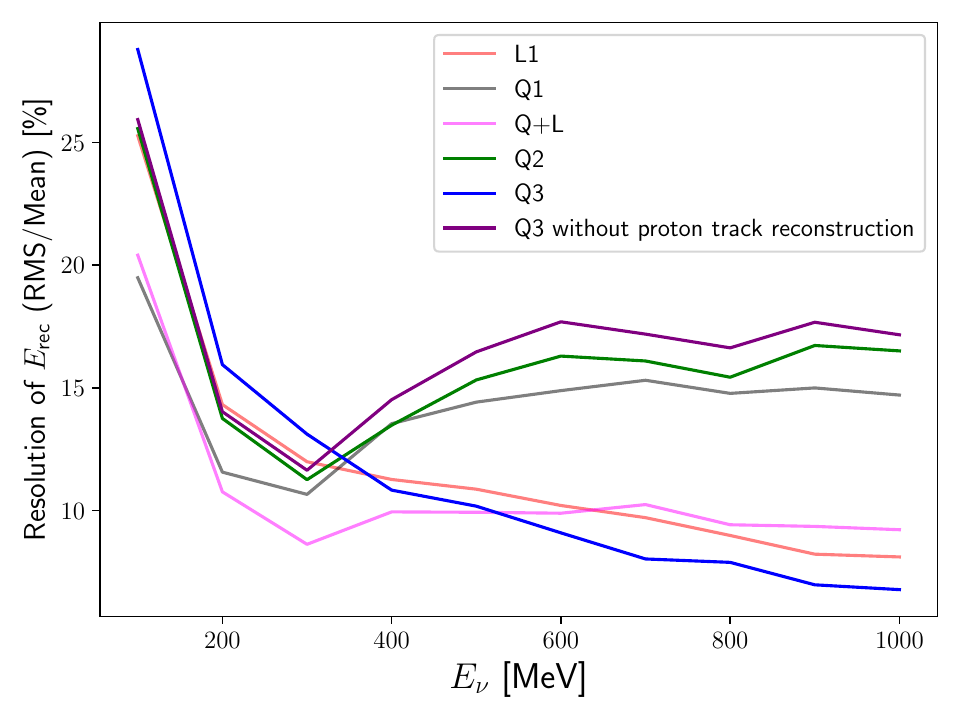}
    \caption{Resolution of the modified Q3 method -- where proton tracks are unable to be separated out and have their total $E_\text{dep}$ reconstructed -- as a function of $E_\nu$, compared with those of the other reconstruction methods. Notice how the modified Q3 method has near-identical performance to the Q2 method.}
    \label{fig:modified Q3}
\end{figure}

Q3 does not have this issue, because if proton tracks can be identified and their dE/dx measured, we can exactly account for the amount of recombination that occurs for each proton track to reliably reconstruct its total $E_\text{dep}$ -- and consequently also its $E_\text{avail}$, since protons deposit almost 100\% of their energy. We can test this reasoning with a modified Q3 reconstruction method in which we assume we are unable to separate out tracks associated with primary protons; instead, we include the charge-deposited energy ($Q_{75}$) of all primary protons and their descendants in the $Q_{75,h}^\text{dots}$ term, to be scaled up by a single factor. As seen in \cref{fig:modified Q3}, this modification to Q3 performs essentially identically to Q2, with degraded resolution at higher $E_\nu$.

We can also compare the sub-GeV reconstruction result with the results for multi-GeV events in Ref.~\cite{Selfcompensatinglight4GeV}. The resolutions we obtained for L1, Q1, and Q3 at 500 MeV and 1000 MeV are generally consistent with Fig. 10 in Ref.~\cite{Selfcompensatinglight4GeV}. However, the resolutions we obtained for Q2 at these two neutrino energies are around 5 percentage points higher (i.e., worse). This is explained by the fact that the scale factor we have employed to scale the hadronic component deposited energy in charge is significantly smaller than the corresponding factor used in the multi-GeV study for reasons discussed in \cref{sec:Ereco_method}.

\section{Sub-GeV Neutrino and Antineutrino Separation}
\label{sec:separation}

As outlined above, the ability to distinguish neutrino events from antineutrino events is crucial for measuring the CP phase in neutrino oscillations. For charged-current interactions, $\nu$ and $\bar{\nu}$ events differ in the sign of the charge of the final state lepton. Previous studies~\cite{mupi_capture} have suggested that $\mu^-$ and $\pi^-$ could be distinguished in an LArTPC by their tendency to capture on Ar nuclei, which does not occur for $\mu^+$ and $\pi^+$. However, no such asymmetry exists between $e^-$ and $e^+$ for $\nu_e$-$\bar{\nu}_e$ separation. And the rarity of charged pions in sub-GeV $\nu_e$ events limits the effectiveness of that avenue.

Another prominent difference between $\nu_e$ and $\bar{\nu}_e$ CC interactions is that for charge to be conserved, $\nu_e$ ($\bar{\nu}_e$) events will almost always involve at least one primary proton (neutron), particularly in the sub-GeV regime where most events are quasi-elastic scattering events. One potential approach for $\nu_e$-$\bar{\nu}_e$ separation could therefore be via final-state topologies, including the use of isolated energy deposits from neutrons~\cite{n_blip_reco_Wan}. However, this is somewhat complicated by final-state interactions, which are highly model-dependent.

With enhancements in light-calorimetry capabilities in LArTPC, $Q$ and $L$ themselves become sufficiently distinct between $\nu_e$ and $\bar{\nu}_e$. Below, we discuss in detail some of the calorimetric features that could be used for charge separation. 

\subsection{Calorimetric Differences between Electron Neutrino and Antineutrino Events in LAr}
\label{sec:calochargeseparation}

\begin{figure}
    \centering
    \includegraphics[width=\linewidth]{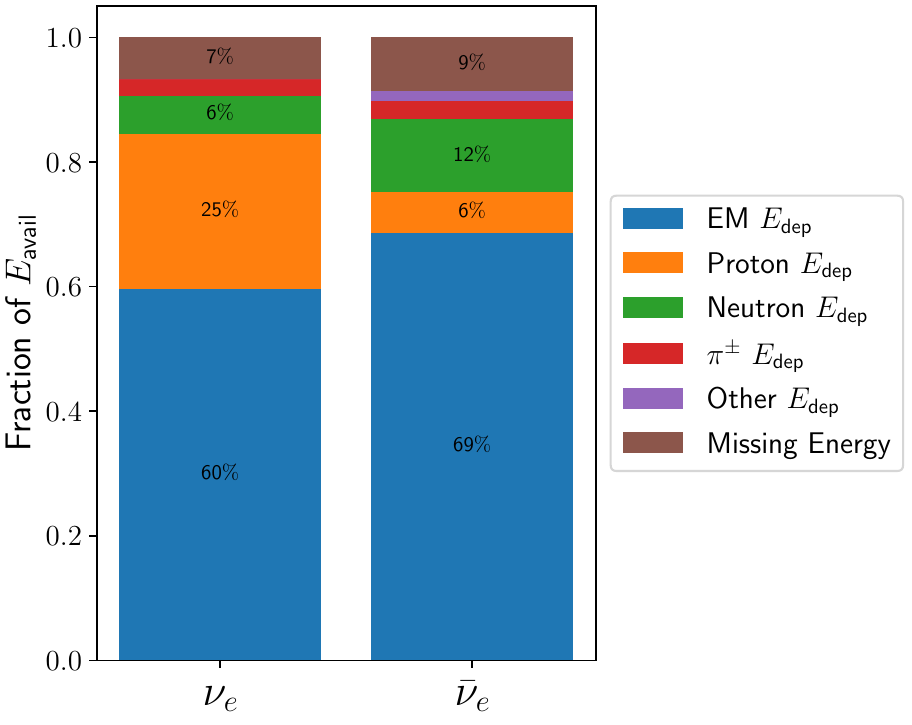}
    \caption{Fractions of total available energy $E_\text{avail}$ distributed among and deposited by final state particles on average across all 10$^4$ simulated sub-GeV (discrete-$E_\nu$) $\nu_e$ and $\antinue$ events each. Note that $\antinue$ events have a much smaller fraction of $E_\text{avail}$ deposited in protons.}
    \label{fig:edep_breakdown_flavor}
\end{figure}

As shown in \cref{fig:edep_breakdown_flavor}, the fraction of $E_\text{avail}$ deposited by primary protons and their descendants across all sub-GeV $\antinue$ events is about a quarter of that for sub-GeV $\nue$ events (6\% versus 25\%). The difference is made up in sub-GeV $\antinue$ events by roughly double the fraction of $E_\text{avail}$ deposited by neutrons and their descendants (11.8\% versus 6.1\%), a larger fraction deposited in the EM component (68.7\% versus 59.6\%), and, as expected, slightly more missing energy (8.6\% versus 6.6\%).

\begin{figure}
    \centering
    \includegraphics[width=\linewidth]{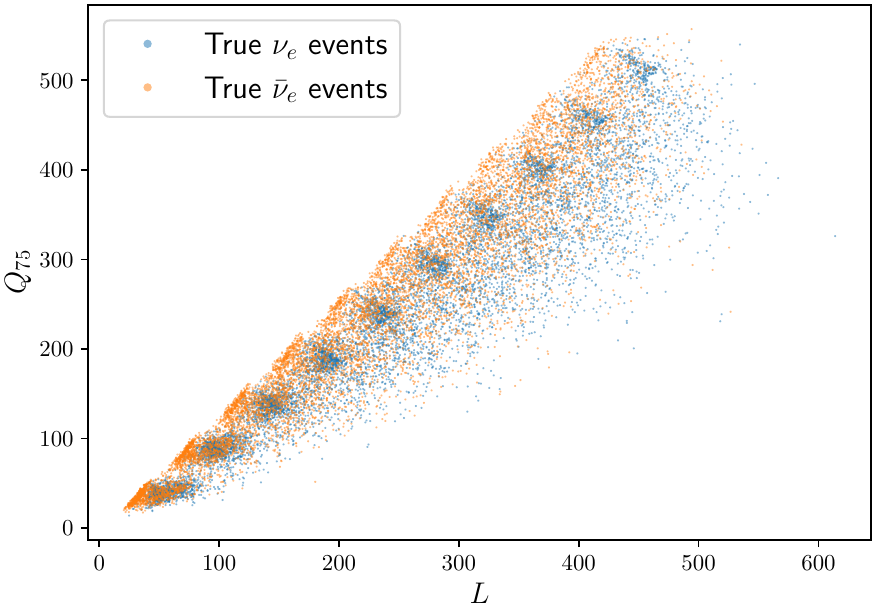}
    \caption{A scatterplot of $Q_{75}$ (i.e $Q$ at 75 keV threshold) vs $L$ for all ten $10^3$-event samples of constant-$E_\nu$ (ranging from 100 MeV to 1000 MeV) of both $\nue$ and $\antinue$. Notice the "strip-like" clusters of events: for each $E_\nu$, the $\nue$ and $\antinue$ clusters are situated at about the same $Q_{75}$ (height), but the $\nue$ cluster is at the larger $L$ (further right).
    These are mostly quasi-elastic scattering events without FSI, whose final states only contain either $1e^-,1p$ for $\nue$ or $1e^+,1n$ for $\antinue$. As such, they most cleanly demonstrate the differences between $\nue$ and $\antinue$ events caused by their involving a proton vs a neutron, respectively. FSI can introduce primary protons in $\antinue$ events / re-absorb primary protons in $\nue$ events, and other interaction processes may have different primary hadrons.}
    \label{fig:qvl_discrete}
\end{figure}

Recall that protons have larger $dE/dx$ and consequently smaller $R_c$ -- particularly at lower energy -- as compared to neutrons and $e^-/e^+$ (which consistently have $R_c\approx0.585$ in this energy regime), which means a proton depositing the same total amount of energy as a neutron or $e^-/e^+$ would result in less additional $Q$ but more $L$. So, the fact that a comparatively much larger share of the $\nue$ $E_\text{avail}$ is deposited by protons suggests that a $\nue$ event of similar $E_\text{avail}$ to an $\antinue$ event will have substantially greater $L$. This is in fact what we observe when we plot $Q_{75}$ against $L$ for the simulated $\nue$ and $\antinue$ event samples with $E_\nu$ from 100 MeV to 1000 MeV (\cref{fig:qvl_discrete}). The $\nue$ and $\antinue$ events appear to be linearly separable. This motivates the idea that a relatively simple machine-learning model trained on these calorimetric features could classify events as $\nue$ or $\antinue$ with reasonable accuracy.

\begin{figure}[h!]
    \centering
    \includegraphics[width=\linewidth]{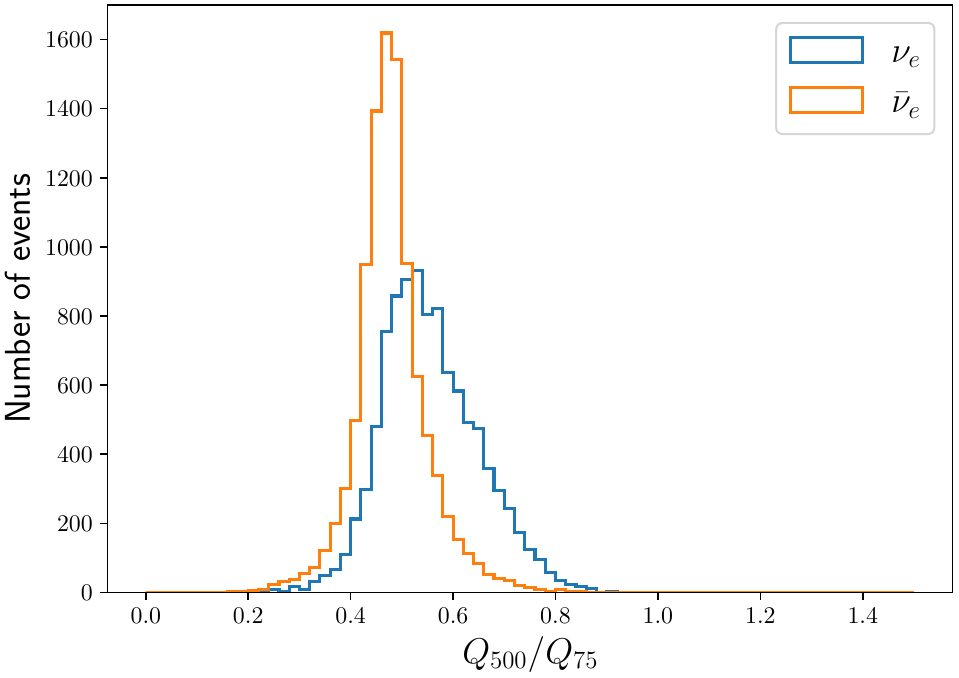}
    \caption{Distributions of the ratio $Q_{500}/Q_{75}$ across both the $\nue$ and $\antinue$ discrete-$E_\nu$ event samples ($10^4$ events each), where $Q_{75}$ and $Q_{500}$ are the total energy-in-ionization-charge recorded for an event given that a threshold of 75 keV and 500 keV respectively has been placed on each individual $Q$-deposit for it to be counted. The bin width is 0.02.}
    \label{fig:q_thres_hist}
\end{figure}

Another calorimetric difference of interest identified between $\nue$ and $\antinue$ events is an apparent asymmetry in the impact of the charge threshold. Specifically, it was observed that, when the threshold is increased from 75 keV to 500 keV in the simulation for individual $Q$-deposits, the relative drop in total $Q$ was on average larger for $\antinue$ events than $\nue$ events. This is well illustrated in \cref{fig:q_thres_hist} where a significant separation is visible between the distribution of $Q_{500}/Q_{75}$ in $\nue$ and $\antinue$ events, with the majority of events with $Q_{500}/Q_{75}\geq0.52$ being $\nue$ and the majority below being $\antinue$. The reasons for this are elucidated in Appendix \ref{sec:appQ}, though, in short, this is another reflection of the fact that protons have a significantly larger on-average share of $E_\text{dep}$ in $\nue$ events than in $\antinue$ events. 

Practically, the charge threshold is bounded below by hardware limits, though it could always be artificially raised during event reconstruction. Meaning that if the charge-readout hardware supports a 75 keV charge threshold, performing the $Q_{75}$-$Q_{500}$ comparison for real events would be feasible.

\subsection{Multivariate Analysis Using Calorimetric Information for Electron Neutrino and Antineutrino Separation}

We have identified a few calorimetric variables above for which we qualitatively observe some separation between $\nue$ and $\antinue$ events. Now, we employ machine-learning (ML) classification algorithms to quantitatively assess the classification performance achievable with a model based on these variables and to better understand how much discriminative power each variable provides. We will compare the selection performances of ML models which use only $Q_{75}$ (i.e. only basic charge calorimetry), that use $Q_{75}$ alongside the light-calorimetry information $L$ or the alternate-charge-threshold information $Q_{500}$, and finally models which utilize all three together ($Q_{75}$, $Q_{500}$ and $L$).

\begin{figure*}
    \centering
    \includegraphics[width=0.75\linewidth]{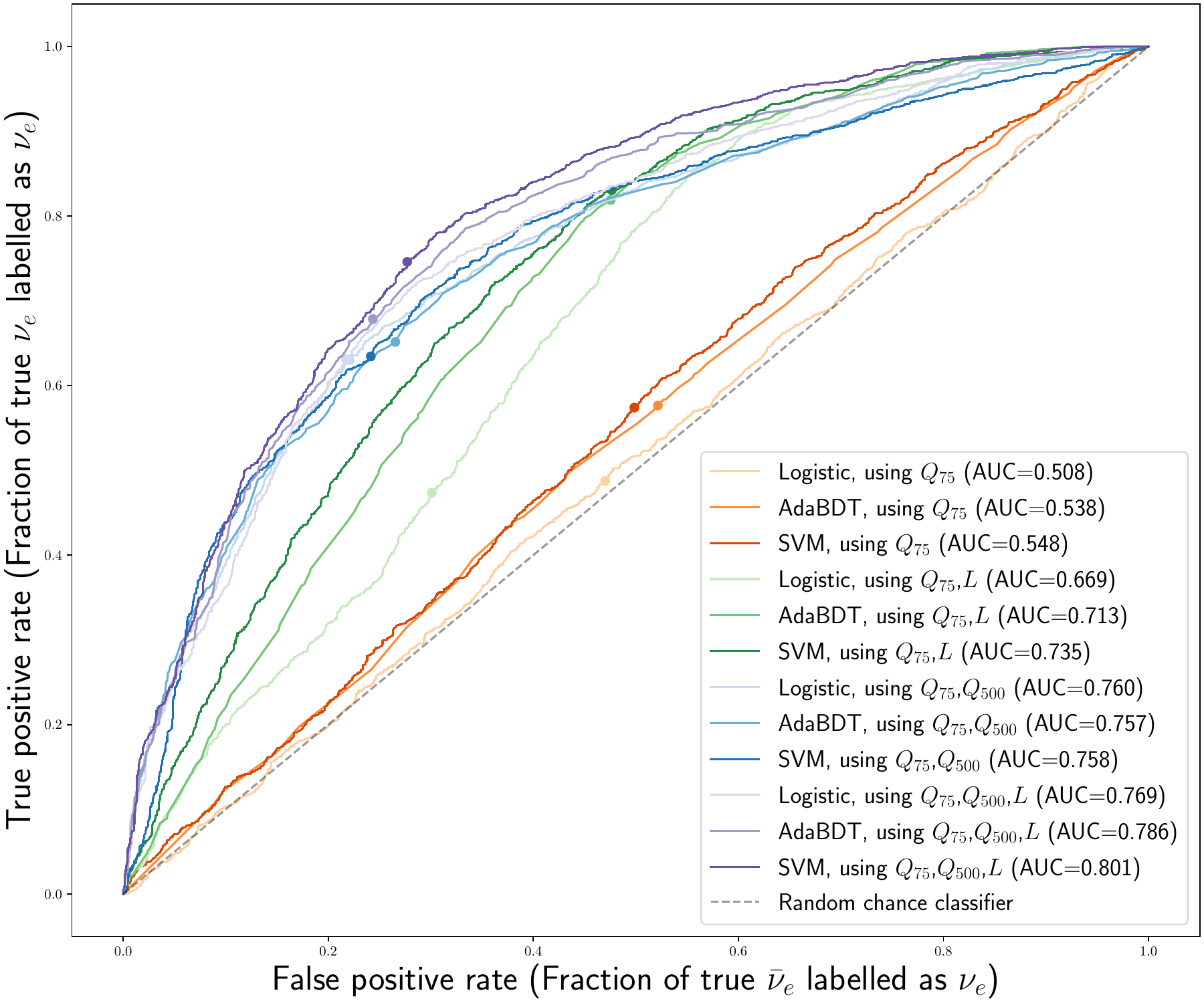}
    \caption{The Receiver Operating Characteristic (ROC) curves of all models computed on the 4000-event discrete-$E_\nu$ evaluation set, using $\nue$ as the positive class and $\antinue$ as the negative. The area under each ROC curve (AUC) is also listed. The dot along each curve is the point corresponding to the default selection threshold or 'cut' used for predictions by the model. 
    For each combination of calorimetric features provided as input, the AdaBDT tends to outperform the Logistic Regression, and the SVM tends to outperform the AdaBDT. This is with the exception of the $Q_{75}; Q_{500}$ models, where the Logit technically achieves the highest AUC, but the three curves overlap to the extent where none can be definitively judged the highest-performing.}
    \label{fig:all_ROC}
\end{figure*}

Since we do not know a priori which specific model architecture best relates the calorimetric variables of interest to the neutrino charge, we test three common models: Logistic Regression (Logit), Support Vector Machine (SVM) with the Radial Basis Function (RBF) kernel, and Adaptive-Boosted Decision Trees (AdaBDT), as implemented in the scikit-learn~\cite{scikit-learn} Python library. All three have hyperparameters that can be freely tuned and generally control the model's complexity. For SVM and Logit, the hyperparameter $C$ is inversely proportional to the strength of the regularization penalty term. RBF SVM also has the $\gamma$ hyperparameter, which controls how smooth or jagged the non-linear decision boundary is. The BDT's complexity is determined by the number of trees in the ensemble and the maximum allowed depth of each individual tree.

To avoid overfitting, we follow a common procedure by splitting the sample into a training set of 16,000 events and an evaluation set of 4,000 events. We ensure that the uniform distribution of $E_\nu$ and equal frequency of $\nue$ and $\antinue$ events are maintained in both. 

Next, to tune the hyperparameters, we search over a grid of possible hyperparameter values for each model and, for each combination, compute the average accuracy across a 10-fold stratified cross-validation. That is, we randomly shuffle and split the development dataset into 10 subsets (again maintaining the uniform class and $E_\nu$ distribution), and for each subset compute the accuracy the model achieves on it upon being trained on the other nine (the accuracy being the fraction of \textit{all} test-events correctly classified). If, for a given combination of hyperparameters, the model performs well across all 10 randomized train-test splits, we can be reasonably confident that the model's performance does not depend on patterns in the sample's noise. Therefore, we select the hyperparameter values that maximize this cross-validated average accuracy. We made sure to expand the search range until none of the best hyperparameters were found to be at the edge of the search grid (except in the 1-feature case with Logit and AdaBDT).

Each model is then re-trained on the entire development set with the optimal hyperparameters, and finally applied to the evaluation dataset.

\subsubsection*{Results}

A clear and comprehensive comparison of the different models across combinations of the calorimetric variables can be achieved by plotting the Receiver Operating Characteristic (ROC) curve for each model, as shown in \cref{fig:all_ROC}. The ROC curve plots the true positive rate (TPR), the fraction of actual $\nue$ events labeled correctly as $\nue$, against the false positive rate (FPR), the fraction of actual $\antinue$ events labeled incorrectly as $\nue$, obtained as the selection threshold is varied across the entire range of scores computed by the model for the evaluation-set events. Note that the TPR and FPR are really the $\nue$ selection efficiency and one minus the $\antinue$ selection efficiency, respectively. Also, note that both the $\nue$ and $\antinue$ selection purities increase with higher TPR (when FPR is fixed) and decrease with higher FPR (when TPR is fixed). An effective classifier will minimize the FPR for each TPR, i.e., reliably reject $\antinue$ even as it selects more $\nue$ (by lowering the selection threshold). 

In \cref{fig:all_ROC}, we find that for the majority of TPR values, the BDT achieves a smaller FPR than the corresponding Logit model, and the corresponding SVM in turn achieves a smaller FPR than the BDT. Meaning that, in general, the SVM is most capable of exploiting calorimetric variables to separate $\nue$ and $\antinue$. This can also be seen by comparing the \textit{areas} under the ROC curves (AUC), which steadily increase going from Logit to BDT to SVM. As such, since we are interested in comparing the discrimination power of different sets of features, we restrict our focus to the SVM models.

\begin{figure}
    \centering
    \includegraphics[width=1\linewidth]{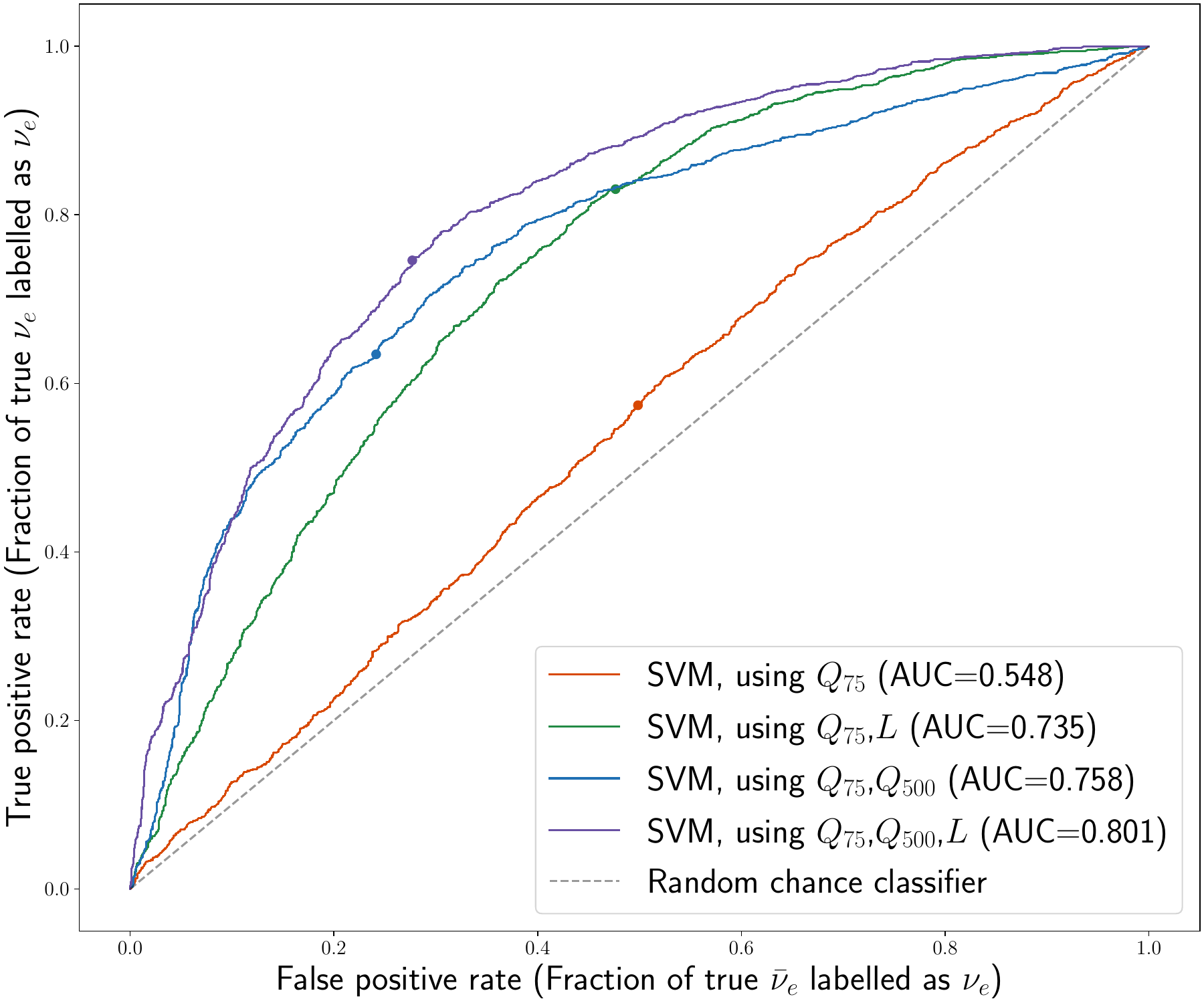}
    \caption{The Receiver Operating Characteristic (ROC) curves obtained on the 4000-event discrete-$E_\nu$ evaluation set for SVM models using different sets of calorimetric variables as predictors. $\nue$ is considered the positive class and $\antinue$ the negative. The area under each curve (AUC) is also given. The dot along each curve corresponds to the model's default selection threshold, as reported in the confusion matrices below. In general, we observe that the more calorimetric information is provided, the better the model performs: for each TPR value, both 2-feature models achieve a much lower FPR than the $Q_{75}$-only model, and the 3-feature model (almost always) achieves an even lower FPR.
    }
    \label{fig:svm_ROC}
\end{figure}

\begin{figure}
    \centering
    \includegraphics[width=1\linewidth]{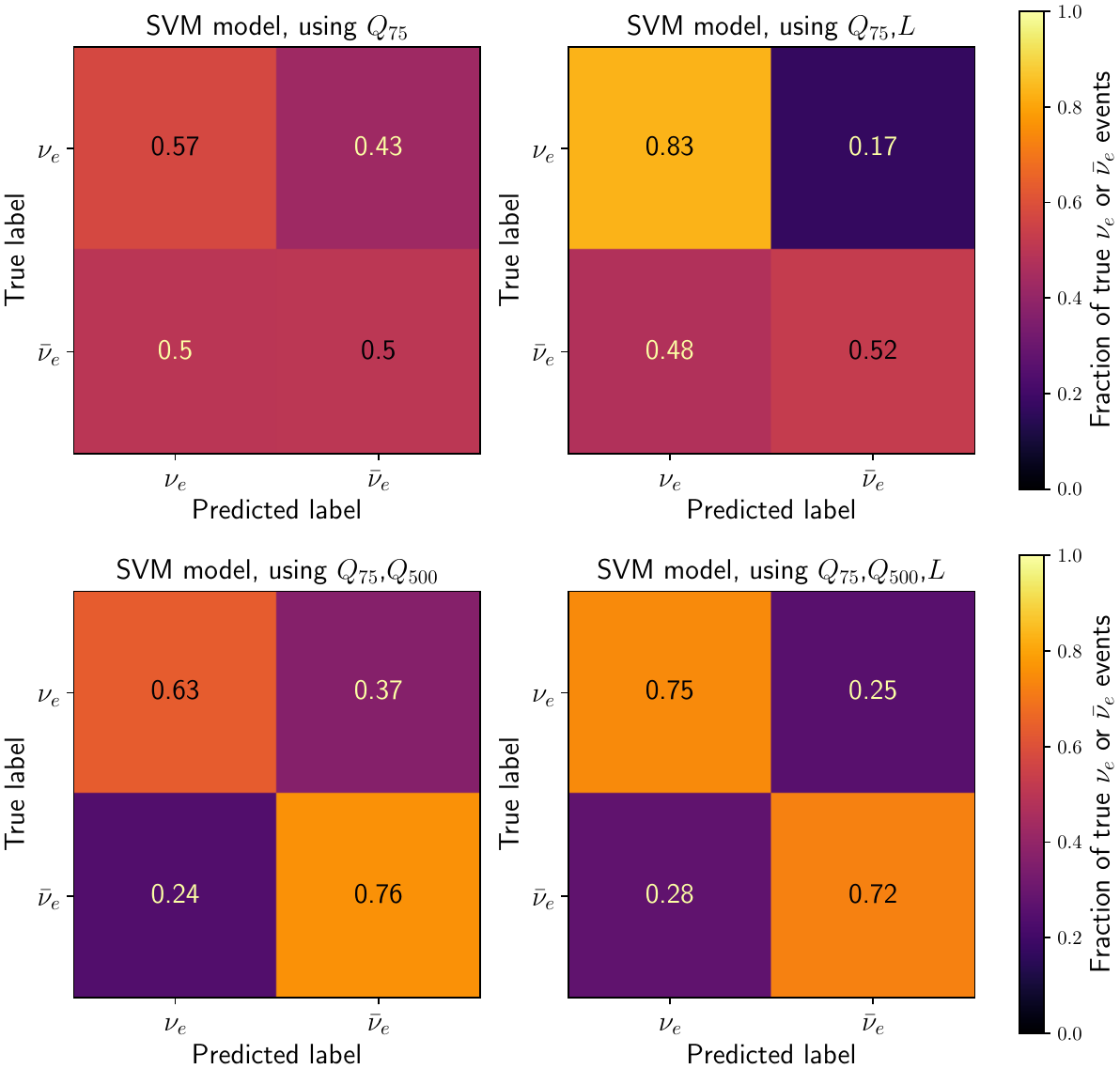}
    \caption{These confusion matrices show how the SVM models split the sets of true $\nue$ and true $\antinue$ events -- 2000 events each -- into predicted $\nue$ and $\antinue$ events. The ideal classifier confusion matrix would have both diagonal entries (which are the efficiencies) be 1 (i.e., 100\% of true $\nue$ ($\antinue$) identified correctly as $\nue$ ($\antinue$)) and 0 in both off-diagonal entries (0\% of true $\nue$ ($\antinue$) identified incorrectly as $\antinue$ ($\nue$)).}
    \label{fig:svm_cm}
\end{figure}

The ROC curves for the SVM models alone are reproduced in \cref{fig:svm_ROC}, along with confusion matrices for the models' predictions at the default selection threshold (\cref{fig:svm_cm}).

We see that relying solely on a single charge-calorimetric variable, even at a low charge threshold, cannot provide any significant discrimination power. The ROC curve of the $Q_{75}$-only SVM model is by far the closest to the diagonal, i.e., to a random-guesser, with an outstandingly low AUC of 0.55 (a random-guesser would have an AUC of 0.50). From the confusion matrix, we find that its classification of $\antinue$ events is essentially no better than random guessing, too. However, it does classify $\nue$ events slightly better than a random guess, raising both $\nue$ and $\antinue$ purities above 50\%.

The addition of the light-calorimetry variable $L$ alongside $Q_{75}$ results in an immediate boost in performance, with the AUC jumping to 0.74. Looking at the confusion matrix, the gain would appear to be largely isolated to the $\nue$ events, with the $\nue$ efficiency rising from 57\% to 83\% while the $\antinue$ efficiency only rises to 52\%. However, the ROC curve makes it clear that by raising the selection threshold, we could instead isolate the gain to $\antinue$: to keep $\nue$ efficiency at 57\% while raising $\antinue$ efficiency to roughly 75\%.

If in place of $L$ we instead used another charge-calorimetry variable taken at a higher charge-threshold $Q_{500}$, we would still obtain a similar boost in performance, reaching an AUC of 0.76. The confusion matrix at the default cut would imply that the $Q_{75}; Q_{500}$ model's performance is more 'balanced' than the $Q_{75}; L$ model: achieving a 76\% $\antinue$ efficiency while still managing 63\% $\nue$ efficiency. Though this is a somewhat unfair comparison: the ROC curves show that, with a higher selection threshold, the $Q_{75}; L$ model could still reach $\sim$70\% $\antinue$ efficiency at the same 63\% $\nue$ efficiency.

One question which naturally arises from the above discussions is: if the separation between $\nue$ and $\antinue$ seen in the relationships of $Q_{75}~\&~L$ and $Q_{75}~\&~Q_{500}$ reflects in both cases the events' differing share of $E_\text{avail}$ in protons, then is the additional information provided by light-calorimetry $L$ redundant with the information provided by charge-calorimetry just with a higher threshold $Q_{500}$? 

It is true that the $Q_{75}; Q_{500}$ model achieves a slightly higher AUC than the $Q_{75}; L$ (by 0.023). However, inspecting the complete ROC curves reveals the full picture: if the priority is to achieve the highest $\nue$ efficiency or TPR at the cost of poor $\antinue$ efficiency and $\nue$ purity, then the $Q_{75}; L$ model with a sufficiently low selection threshold is superior. Whereas, if the priority is to minimize the FPR -- and so maximize the $\nue$ purity and $\antinue$ efficiency -- at the expense of the $\nue$ efficiency, then the $Q_{75}; Q_{500}$ model with a high selection threshold is the better option. Moreover, the ROC curves do in fact intersect, meaning for a certain (non-trivial) choice of two thresholds the two models actually would perform identically well (with an 84\% $\nue$ efficiency but essentially-50\% $\antinue$ efficiency).

Were $L$ to be redundant with $Q_{500}$ for charge separation, we would see no significant gain in performance from the $Q_{75}; Q_{500}$ model to the model which uses all three: $Q_{75}; Q_{500}; L$. However this is refuted by the default-threshold confusion matrices: even though the $Q_{75}; Q_{500}; L$ model sees a slightly lower $\antinue$ efficiency (72\%, down from 76\%), the increase in $\nue$ efficiency -- from 63\% to 75\% -- is enough that the overall accuracy rises from 69.7\% to 73.5\% (the highest accuracy achieved among all models tested), and both purities increase as well. In fact, out of all models evaluated, only the SVM using all three calorimetric variables scores above 70\% for all four efficiency and purity metrics at the default cut.

The superior performance attainable upon the inclusion of $L$ is validated by comparing the full ROC curves. The $Q_{75}; Q_{500}; L$ model's curve clearly surrounds the other three -- for almost any value of TPR, i.e., $\nue$ efficiency (except around 0.3-0.45), it provides better $\antinue$ efficiency (lower FPR) and thus $\nue$ purity than the other three feature sets. One sees that by choosing lower selection cuts, one could improve the $\nue$ efficiency above the default-threshold $Q_{75}; Q_{500}$ model without \textit{any} loss of $\antinue$ efficiency or vice versa. The $Q_{75}; Q_{500}; L$ model has the highest AUC of 0.80 (0.043 above the $Q_{75}; Q_{500}$ model's). 

It is thus clear that, in order to achieve the optimal overall classification performance, all three calorimetric variables -- the two charge-threshold $Q$ values \textit{and} the light-calorimetry $L$ -- are necessary. 

We have shown that even a model as simple as SVM is able to achieve substantial $\nue$-$\antinue$ classification performance with only three calorimetric variables. Naturally, a practical machine-learning neutrino flavor-reconstruction model would be far more complex and would include a broader array of (less model-dependent) event information, such as track lengths, the number of blips, etc., thereby yielding stronger, more robust classification. Nevertheless, this study emphasizes the benefits of advances in calorimetry capabilities, both in further developing light calorimetry and in lowering the charge threshold to enable the $Q_{75}$-$Q_{500}$ comparison to be exploited. 

\section{Sub-GeV Neutrino Direction Reconstruction}
\label{sec:dir}

\begin{figure}[htbp]
    \centering
    \includegraphics[width=1.0\linewidth]{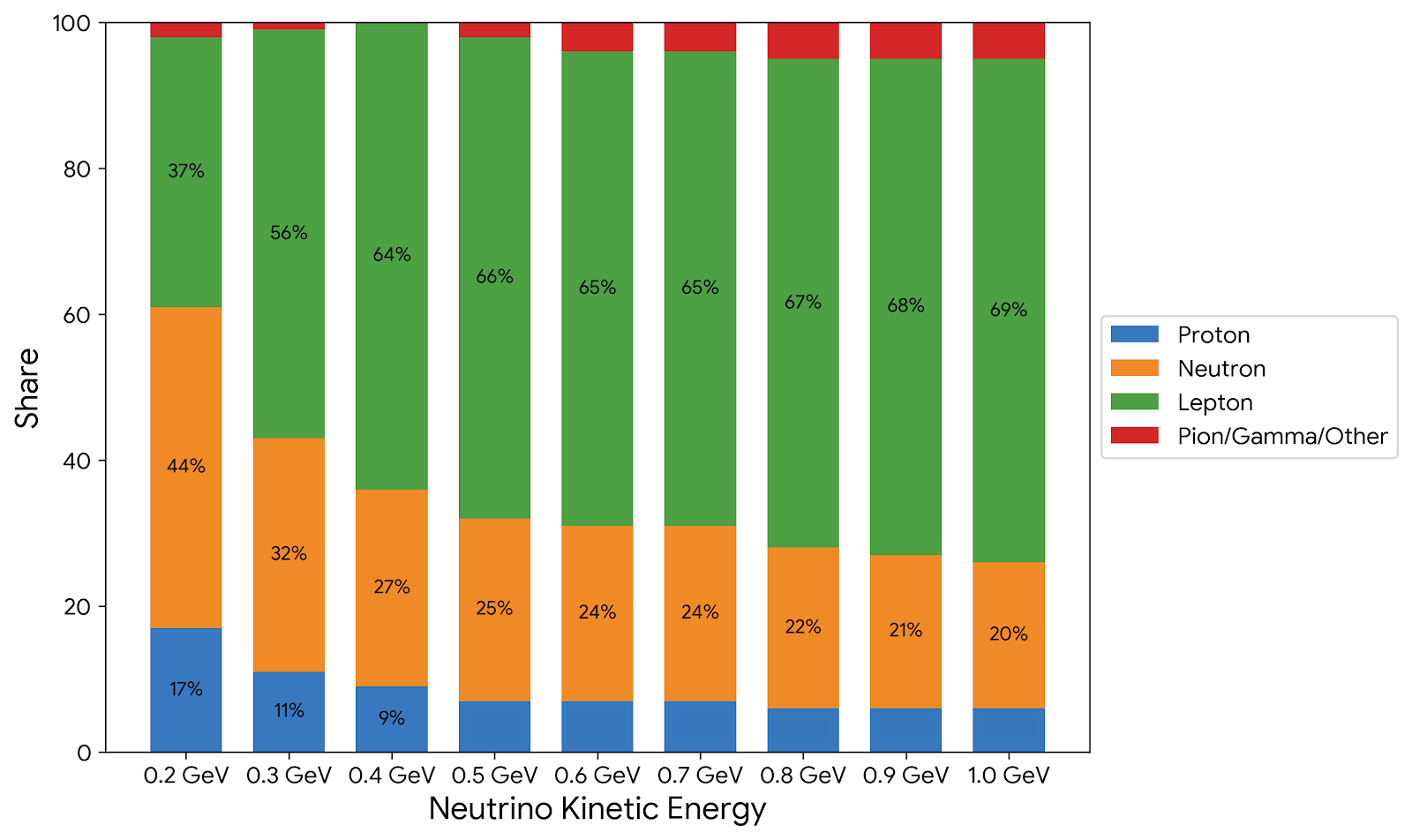}
    \caption{Stacked distribution of the kinetic-energy fraction carried by each final-state particle emerging from the $\bar{\nu}_\mu$ interaction vertex at discrete neutrino energies.}
    \label{fig:stack_plot}
\end{figure}

In addition to neutrino energy reconstruction and charge separation, accurate neutrino direction reconstruction is also crucial for precise measurements of neutrino oscillations using sub-GeV atmospheric neutrinos~\cite{subGeV_atmnuosc}. The current atmospheric neutrino direction reconstruction typically employs two paradigms: using lepton-only information or using all reconstructed particles~\cite{DUNE_HD_Atmnureco}. The first method estimates the direction of the incident neutrino solely from the reconstructed primary lepton. Although robust, this ignores the significant momentum contribution of the hadronic system. The second approach attempts to reconstruct the hadronic system but is fundamentally limited by the missing momentum from neutral particles, primarily neutrons. In LArTPCs, neutrons only manifest through sporadic energy deposits~\cite{Friedland_Li_nuEres} or secondary tracks such as protons~\cite{MicroBooNE_neutron_via_2nd_proton}. This is particularly problematic for antineutrino interactions, where the charged-current process frequently ejects one or more neutrons. Neutrons share up to 44\% of the total energy of all outgoing particles, as shown in \cref{fig:stack_plot}. In the sub-GeV region, many protons are also not energetic enough to leave a track, so a large fraction of proton activity is missed. These further motivate neutron reconstruction and low-energy hadronic system reconstruction, which is indispensable for neutrino direction reconstruction.

We present a method to reconstruct the neutron momentum vector in LArTPCs. The discussion will focus on neutron direction reconstruction (\cref{sec:dir:neutron-direction}) and neutron momentum amplitude reconstruction (\cref{sec:dir:neutron-momentum}). After these, we will evaluate how the sub-GeV neutrino reconstruction performance improves with neutron momentum reconstruction (\cref{sec:dir:nu-direction}).

\subsection{Neutron direction reconstruction}
\label{sec:dir:neutron-direction}

Our neutron direction reconstruction methodology employs a proximity-based algorithm in which the incident direction is defined by the vector from the neutrino interaction vertex to the spatially nearest energy deposition. This approach is predicated on the assumption that the first neutron scattering provides the most accurate representation of the primary neutron direction; subsequent interactions are characterized by isotropic scattering, leading to a random walk that degrades directional information. 

\begin{figure}[htbp]
    \centering
    \includegraphics[width=0.9\linewidth]{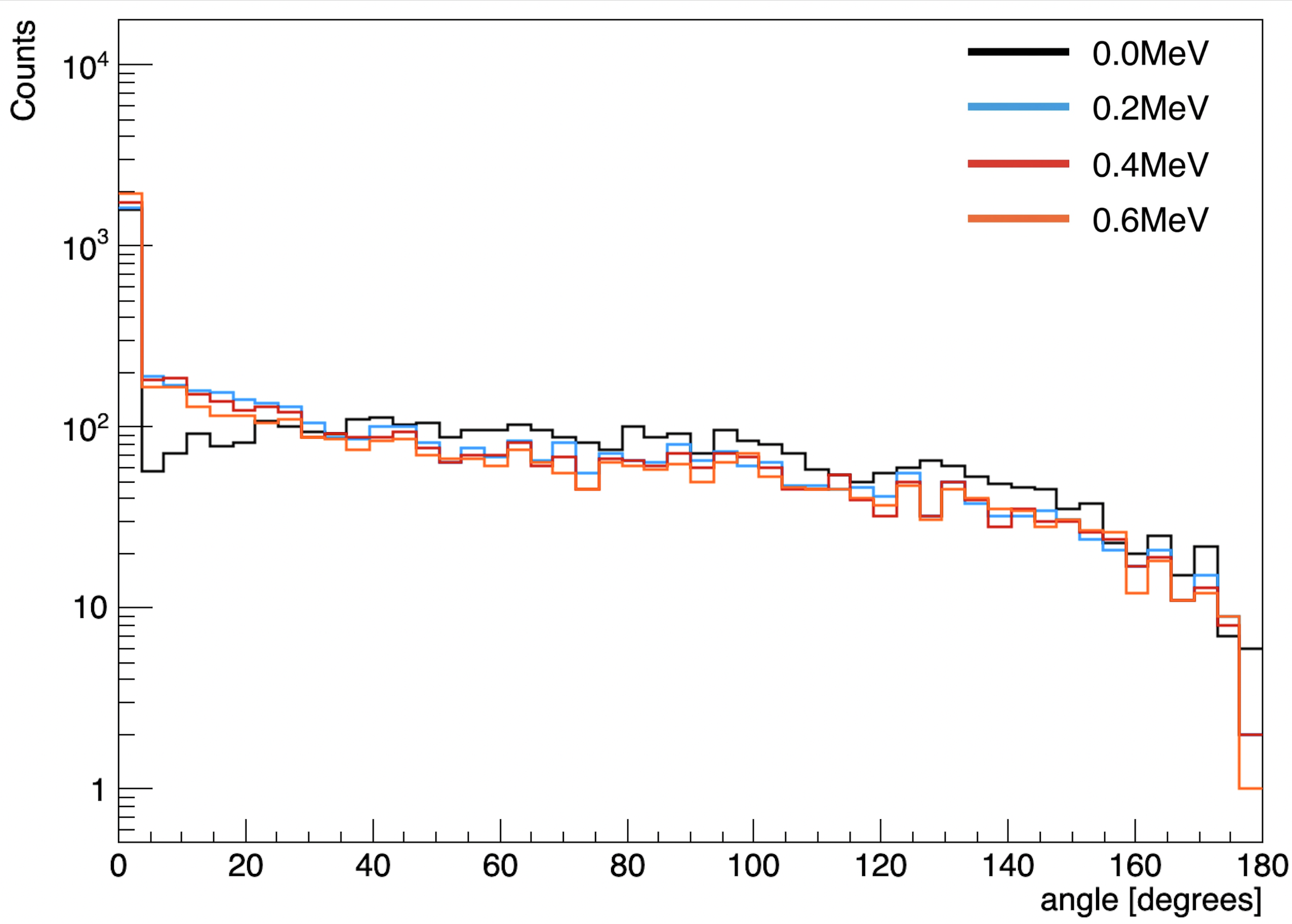}
    \caption{Angular difference between the earliest and nearest neutron-induced energy deposition from events with exactly one neutron from the neutrino vertex. The peak near $0^\circ$ indicates that the nearest deposition coincides with the earliest deposition.}
    \label{fig:earliest_nearest}
\end{figure}

Although the chronologically earliest neutron interaction point would provide the optimal direction reconstruction, the limited temporal resolution of isolated, low-energy deposits in current LArTPCs precludes this possibility. Consequently, the spatially nearest neutron deposition to the neutrino interaction vertex serves as a proxy for the earliest neutron interaction. \cref{fig:earliest_nearest} suggests that this spatial proximity maintains high directional fidelity, as the nearest deposition remains highly correlated with the primary neutron interaction point for a majority of events.

\begin{figure}[t]
\centering
\includegraphics[width=0.48\textwidth]{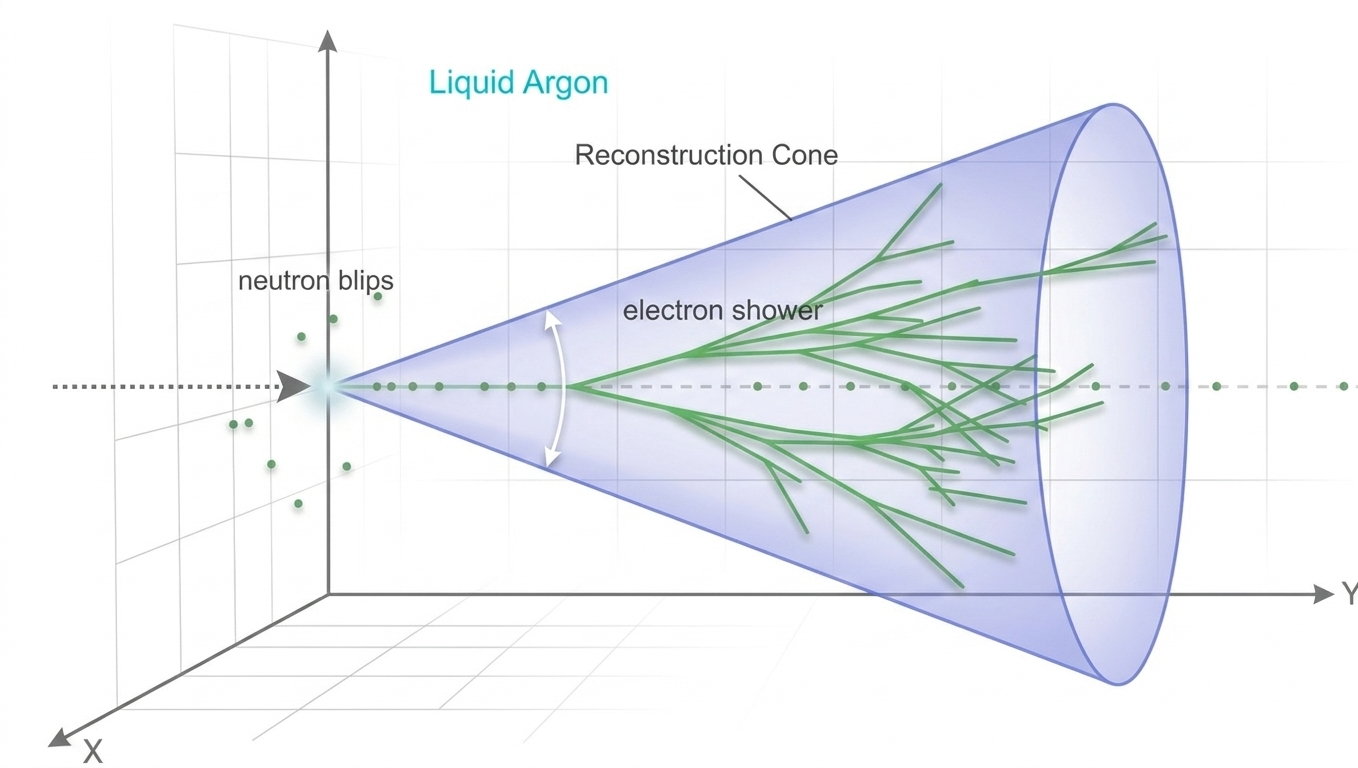}
\caption{Schematic representation of the lepton exclusion cone. The geometric veto is applied to distinguish neutron energy deposits from those coming from primary leptons. This drawing has been produced with the assistance of Google Gemini.}
\label{fig:lepton_cone}
\end{figure}

A critical component of this methodology is discriminating neutron-induced isolated energy deposits from competing backgrounds. Notably, primary electrons that propagate as electromagnetic showers frequently produce fragmented energy signatures that topologically mimic neutron-induced hits. We propose a geometric isolation based on a conical veto volume centered at the primary neutrino vertex as shown in \cref{fig:lepton_cone}. The central axis and longitudinal extent of the cone are determined by the trajectory and range of the primary outgoing lepton, which are reconstructed reliably at this energy range by algorithms used in current LArTPCs~\cite{WireCell, DUNE_pandora_shower_reco}. This geometric isolation is designed to suppress lepton-induced backgrounds by excluding the majority of the volume where the lepton deposits its energy.

A successful neutron direction reconstruction is defined when the nearest energy deposition among all depositions outside the conical veto volume is associated with the true primary neutron. There are some failure modes, for example, when the nearest deposition originates from non-neutron backgrounds, and when the primary neutron undergoes its initial interaction within the excluded conical volume. By scanning the cone opening angle $\theta_{c}$, we optimize $\theta_{c}$ to maximize the efficiency of picking up the nearest neutron energy deposition outside the veto cone ($N_{\text{successful}}$),
\begin{equation}
    \epsilon = \frac{N_{\text{successful}}}{N_{\text{all events}}}.
\end{equation}

\begin{figure}[t]
    \centering
    \includegraphics[width=0.9\linewidth]{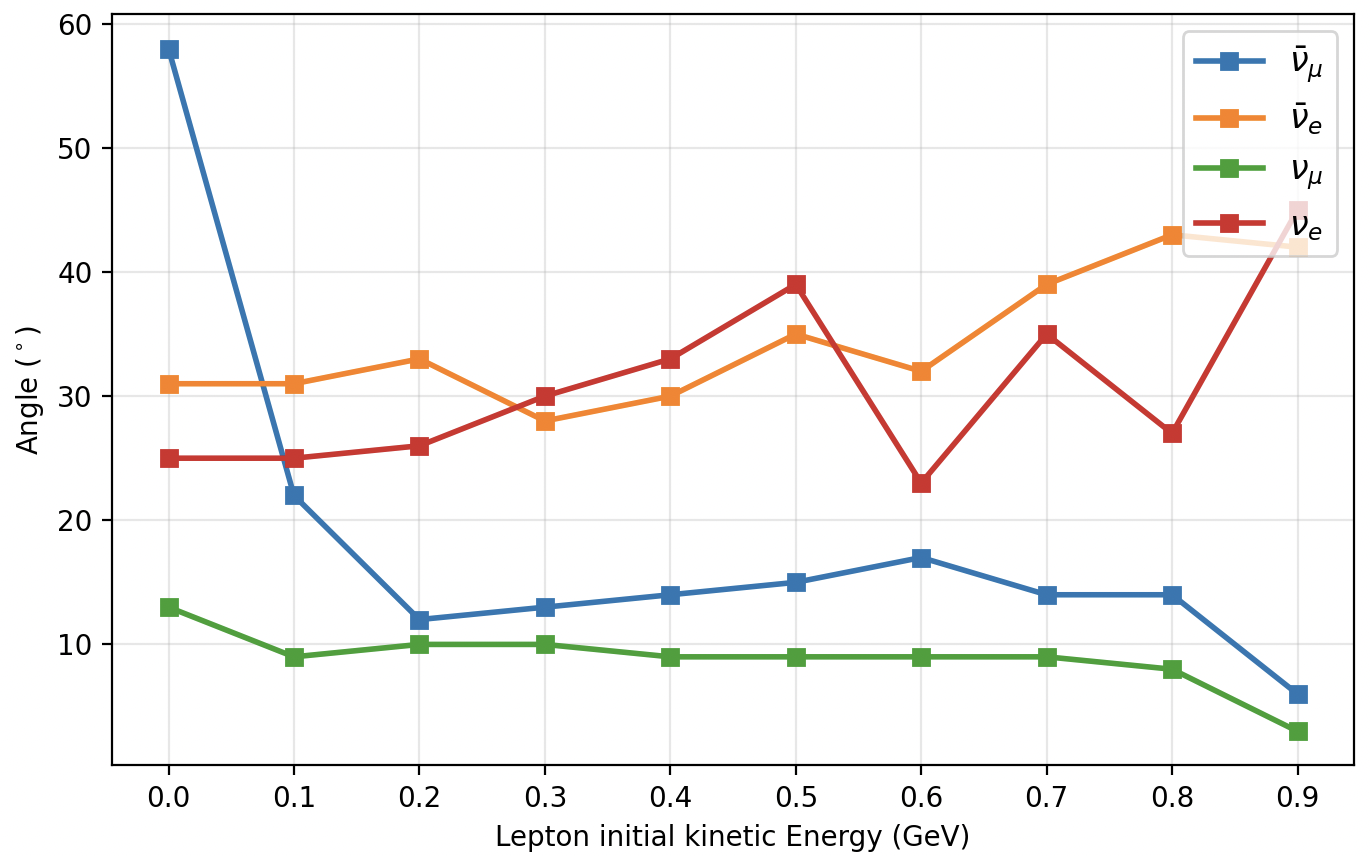}
    \caption{Optimal lepton exclusion cone size, $\theta_{c}$, as a function of lepton energy in four neutrino samples. Each cone size maximizes the efficiency of the selected nearest energy deposit outside the veto cone, which is associated with the true primary neutron.}
    \label{fig:optimalopeningangle}
\end{figure}

The optimal cone geometry is highly dependent on the lepton flavor and its associated energy deposition profile, as shown in \cref{fig:optimalopeningangle}:
\begin{itemize}
    \item \textbf{$\bar{\nu}_{\mu}$ and $\nu_{\mu}$:} In $\bar{\nu}_{\mu}$ interactions, the resulting $\mu^{+}$ consistently undergoes decay-at-rest, producing a Michel electron ($e^{+}$). At low energies, this secondary deposition necessitates a wider cone. The optimal exclusion cone narrows as the muon's kinetic energy increases. 
    
    Conversely, for $\nu_{\mu}$ interactions, approximately 75\% of $\mu^{-}$ in the simulation are captured on the $^{40}$Ar nucleus~\cite{LArIAT_mucapture}. This nuclear capture suppresses the Michel electron signal, thereby allowing a more constrained exclusion cone than in the antineutrino channel.
    
    \item \textbf{$\nu_{e}$ and $\bar{\nu}_{e}$:} The electron or positron typically travels approximately one radiation length ($X_0 \approx 14$~cm in LAr) before initiating an electromagnetic shower. The transverse profile of this shower is characterized by the Molière radius ($R_M \approx 10$~cm)~\cite{LArproperties}. Because $R_M$ is a property of the medium, the shower width remains relatively constant regardless of energy; thus, the optimal cone angle is largely uncorrelated with the lepton's kinetic energy as shown in \cref{fig:optimalopeningangle}.
\end{itemize}
Additionally, we apply a detector energy threshold of 0.6 MeV to simplify the treatment of the dominant $^{39}\text{Ar}$ radiological background in bulk LAr, whose decay produces electrons with a maximum energy of 565 keV. In \cref{sec:appA}, we demonstrate that the $^{39}\text{Ar}$ background has a non-negligible impact on the final result, even if the region of interest for finding the neutron's nearest energy deposit is small. When we lower the energy threshold from 0.6 MeV to 0.3 MeV or 0.1 MeV, the improvement contributed by neutron momentum reconstruction will be diluted in the final reconstruction of the neutrino direction.  

\begin{figure}[H]
    \centering
    \includegraphics[width=0.9\linewidth]{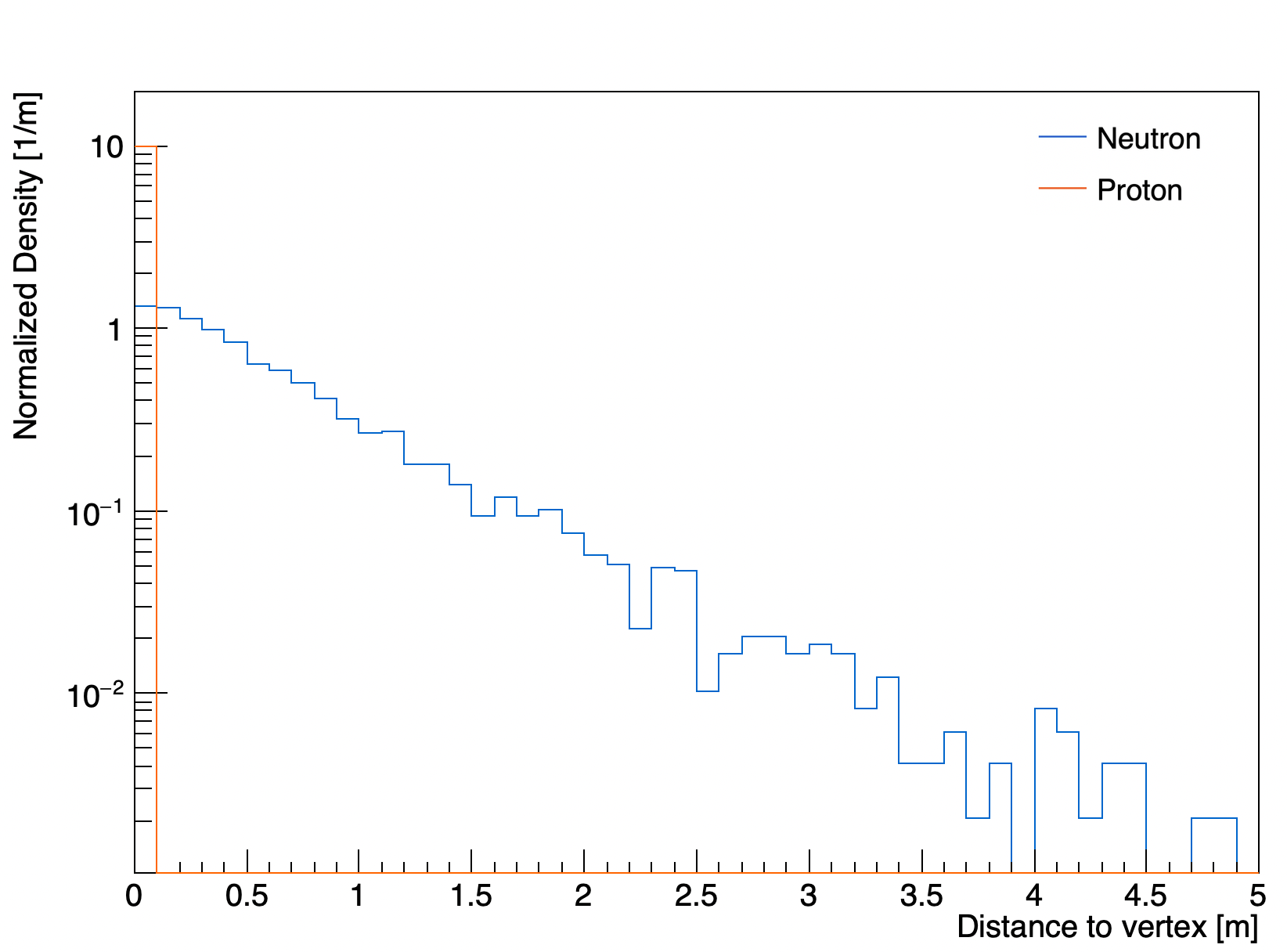}
    \caption{Neutron nearest energy deposition's distance to the neutrino vertex compared with proton energy deposition distance to the neutrino vertex, whose kinetic energy is below 70 MeV. We exclude this noise from the low-energy protons by simply using a veto sphere, which does not affect neutron direction reconstruction.}
    \label{fig:neutron_proton}
\end{figure}

Localized energy deposits from low-energy protons also constitute a background for this analysis. We consider a conservative kinetic energy threshold of 70 MeV for a proton to produce a detectable track in LAr; protons with kinetic energies below this threshold are considered blip-like. Because sub-threshold proton energy depositions are tightly clustered around the neutrino interaction vertex, they can be effectively excluded by applying a 5 cm radius spherical spatial cut centered on the neutrino vertex. Crucially, this spherical cut preserves the majority of the primary neutron energy depositions central to this study, as illustrated in \cref{fig:neutron_proton}.

From the simulation, we also observed that most of the earliest neutron energy deposition is within a 1.5-meter radius from the neutrino vertex. We adopt this region of interest (ROI) for identifying the nearest neutron energy deposition. Using a larger ROI volume would include more noise and affect the reconstruction. We reserve this discussion for the next section where we describe the neutron momentum reconstruction. At the sub-GeV energy range, it is also possible to have multiple neutrons produced as a result of various nuclear effects. Experimentally, it is challenging to resolve these. In these cases, we treat the energy depositions from all neutrons as a single effective low-energy system emanating from the vertex. 

\begin{figure}
    \centering
    \includegraphics[width=0.9\linewidth]{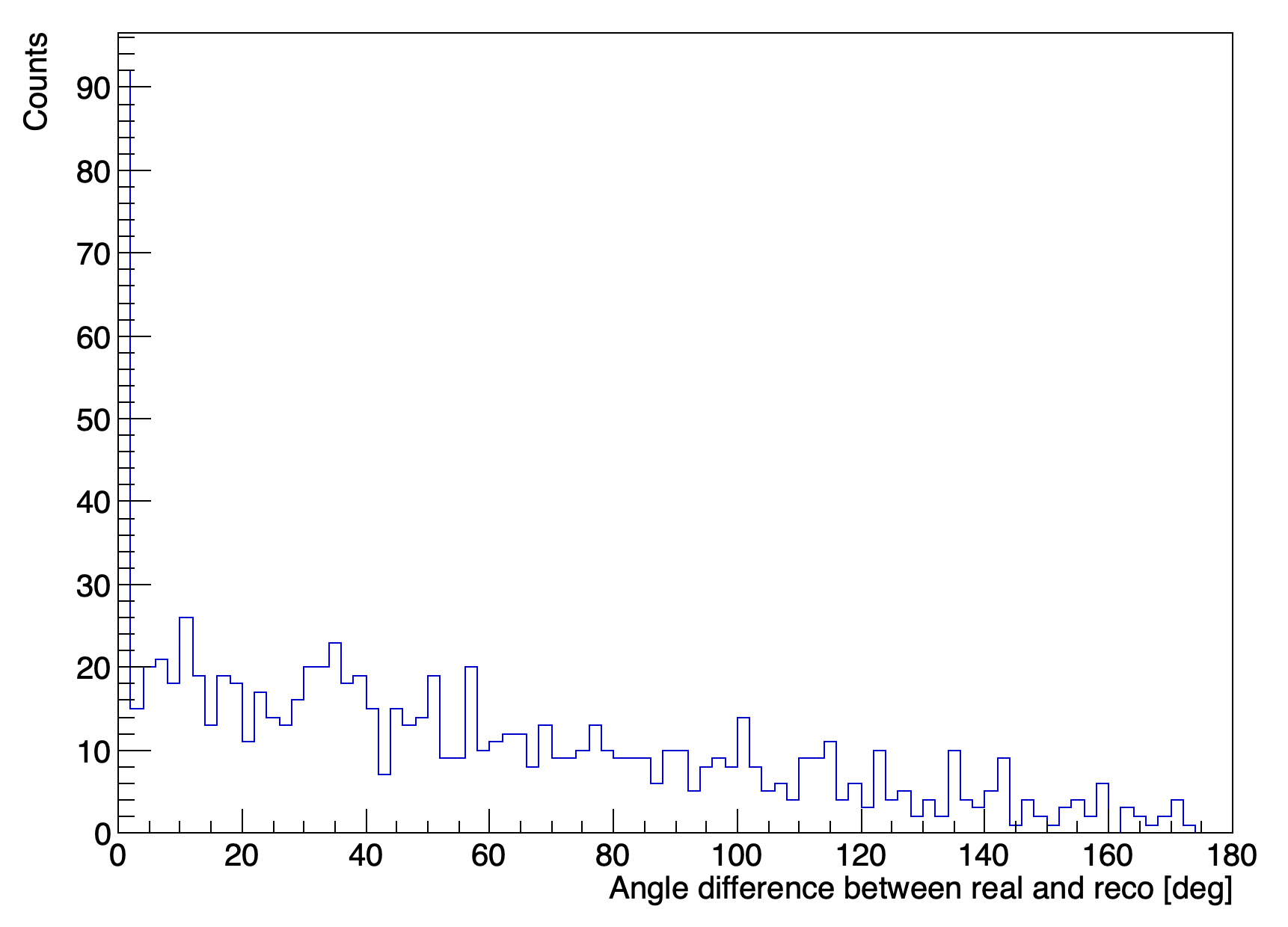}
    \caption{Angle difference between the reconstructed neutron direction and the true neutron momentum direction for 0.5 GeV $\bar{\nu}_{\mu}$ events.}
    \label{fig:neutrondirreco}
\end{figure}

To summarize, for the neutron direction analysis, within a ROI sphere of radius 1.5 m and excluding an inner 5-cm radius spherical veto region around the neutrino vertex, all localized, blip-like energy depositions above 600 keV outside the lepton veto cone are considered neutron energy deposit candidates. The vector pointing from the neutrino vertex to the nearest of these candidate depositions is taken as the reconstructed neutron direction. The performance of the neutron direction reconstruction is shown in \cref{fig:neutrondirreco}. The peak around zero shows those events where the earliest neutron energy deposit is, in fact, the nearest energy deposit, and we perfectly reconstruct these events. Deviations from the true neutron direction exist, but the fraction of these events decreases at larger angle deviations and reaches a minimum in the opposite direction of the neutron's true momentum.

\subsection{Neutron momentum amplitude reconstruction}
\label{sec:dir:neutron-momentum}

\begin{figure*}
    \centering
    \includegraphics[width=0.8\linewidth]{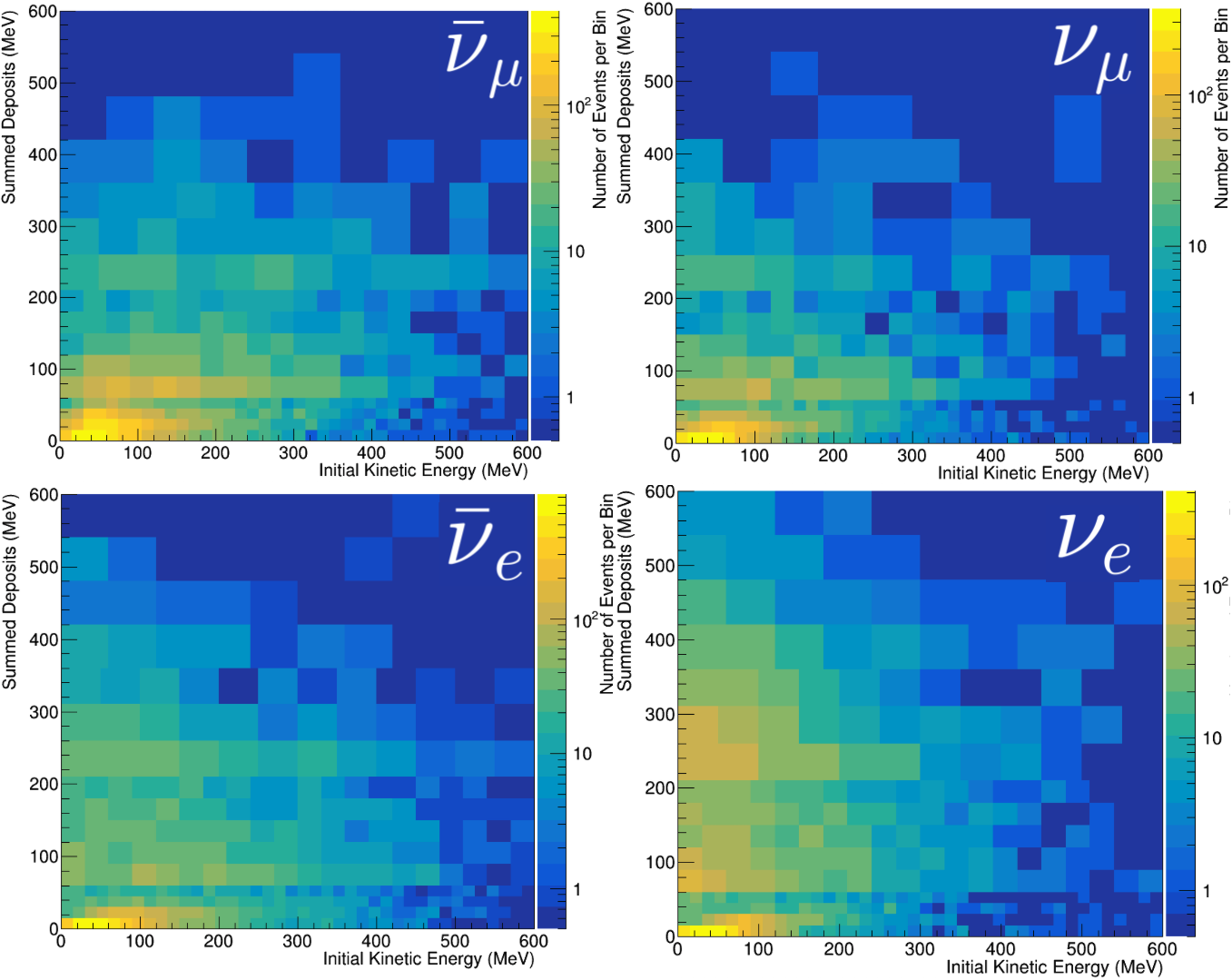}
    \caption{Relationship between the sum of neutron energy deposition within 1.5 m from the neutrino vertex and the true kinetic energy of the neutron.}
    \label{fig:sumedep_truenKE}
\end{figure*}

The neutron momentum amplitude reconstruction is based on the relativistic relation between the neutron kinetic energy ($K_{n}$) and its momentum amplitude ($p_{n}$):
\begin{equation}
p_{n} = \sqrt{K_{n}^2 + 2K_{n}m_{n}}.
\end{equation}
The task is to reconstruct the initial neutron kinetic energy ($K_{n}$) from the energy deposition near the neutrino vertex. The strategy is to sum the candidate neutron energy depositions and find the relation between this quantity and the true neutron kinetic energy. In the simulated sample, about 60\% of the neutron energy deposition above the detection threshold of 0.6 MeV falls into a sphere of 1.5 m radius around the neutrino vertex. To start with, we first exclude depositions more than 1.5 m away from the neutrino vertex, and then use cones as defined in the previous section to exclude all track-like energy depositions, including those from muons, pions, and protons with kinetic energy above 70~MeV. The summed candidate neutron energy deposits and the true neutron kinetic energy are plotted in 2-dimensional histograms for four neutrino flavors in \cref{fig:sumedep_truenKE}. 

We divide the plot into three regions and use different binning schemes to record the distribution, ensuring we do not infer a relationship from insufficient statistics. The bin sizes are 15 MeV $\times$ 15 MeV below 60 MeV of summed energy deposit, 30 MeV $\times$ 30 MeV for summed energy deposit between 60 MeV and 210 MeV, and 50 MeV $\times$ 50 MeV when the summed energy deposit is above 210 MeV, respectively. For each summed-deposits bin, we find the most probable initial kinetic energy and use this as the reconstructed neutron energy. Note that above 360 MeV of summed energy deposition, the number of events in our simulation is limited; we therefore do not extract a most probable initial kinetic energy, and we ignore these events during the reconstruction.

We observe that there are many entries in which the summed neutron deposition exceeds the true kinetic energy, especially in neutrino events, where often no neutron is produced in the interaction. These extra energy depositions are from non-neutron sources and are incorrectly included in the 1.5 m-radius sphere. We also exclude these events in this analysis.

\begin{figure*}
    \centering
    \includegraphics[width=0.8\linewidth]{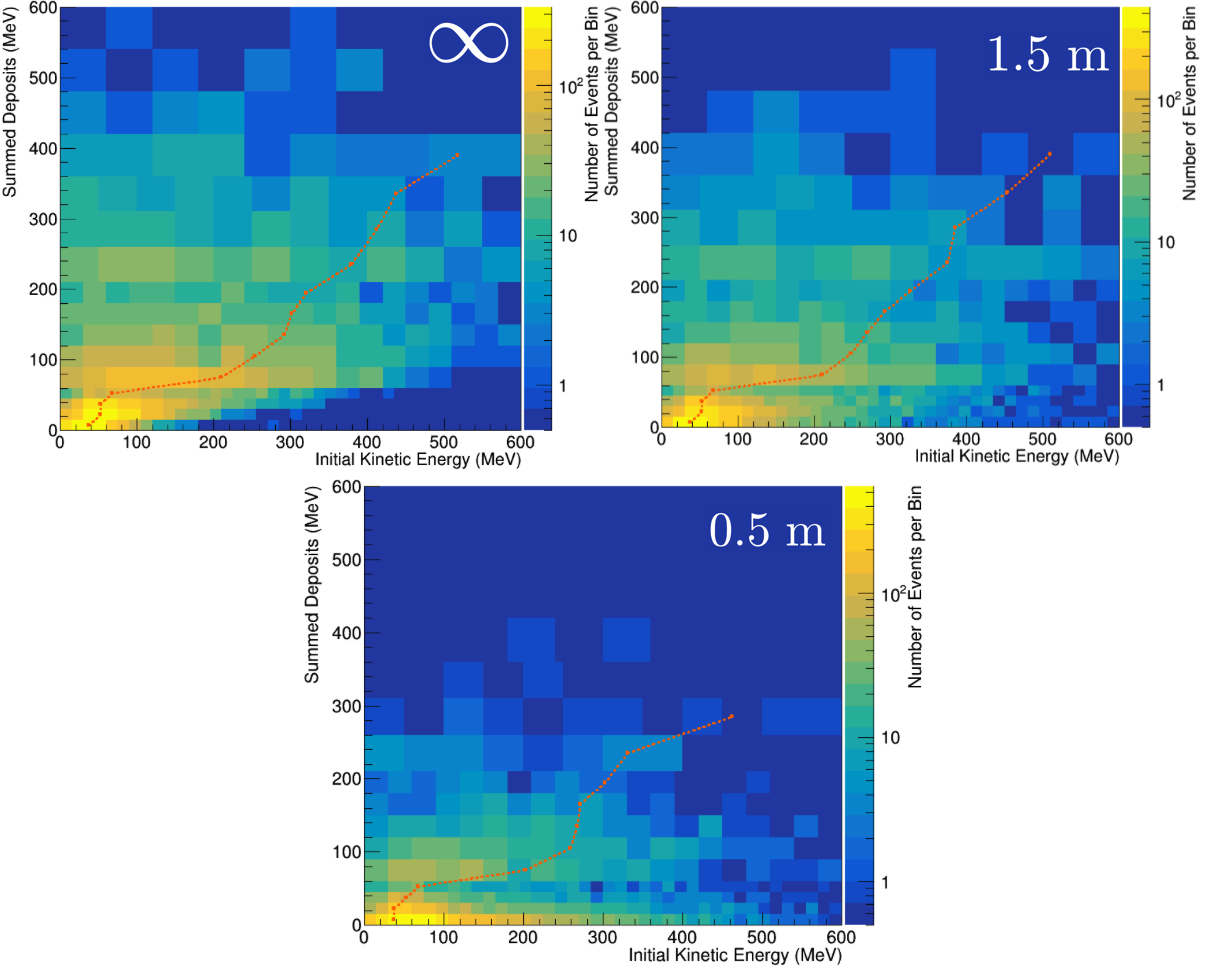}
    \caption{Comparison of the relationship between neutron kinetic energy and summed neutron energy deposit candidate in different ROI volume sizes: an infinite volume (top left), a sphere of 1.5-meter (top right), and a 0.5-meter radius around the neutrino vertex.}
    \label{fig:neutronEcorrelation_radiusscan}
\end{figure*}

\begin{figure}
    \centering
    \includegraphics[width=1.0
    \linewidth]{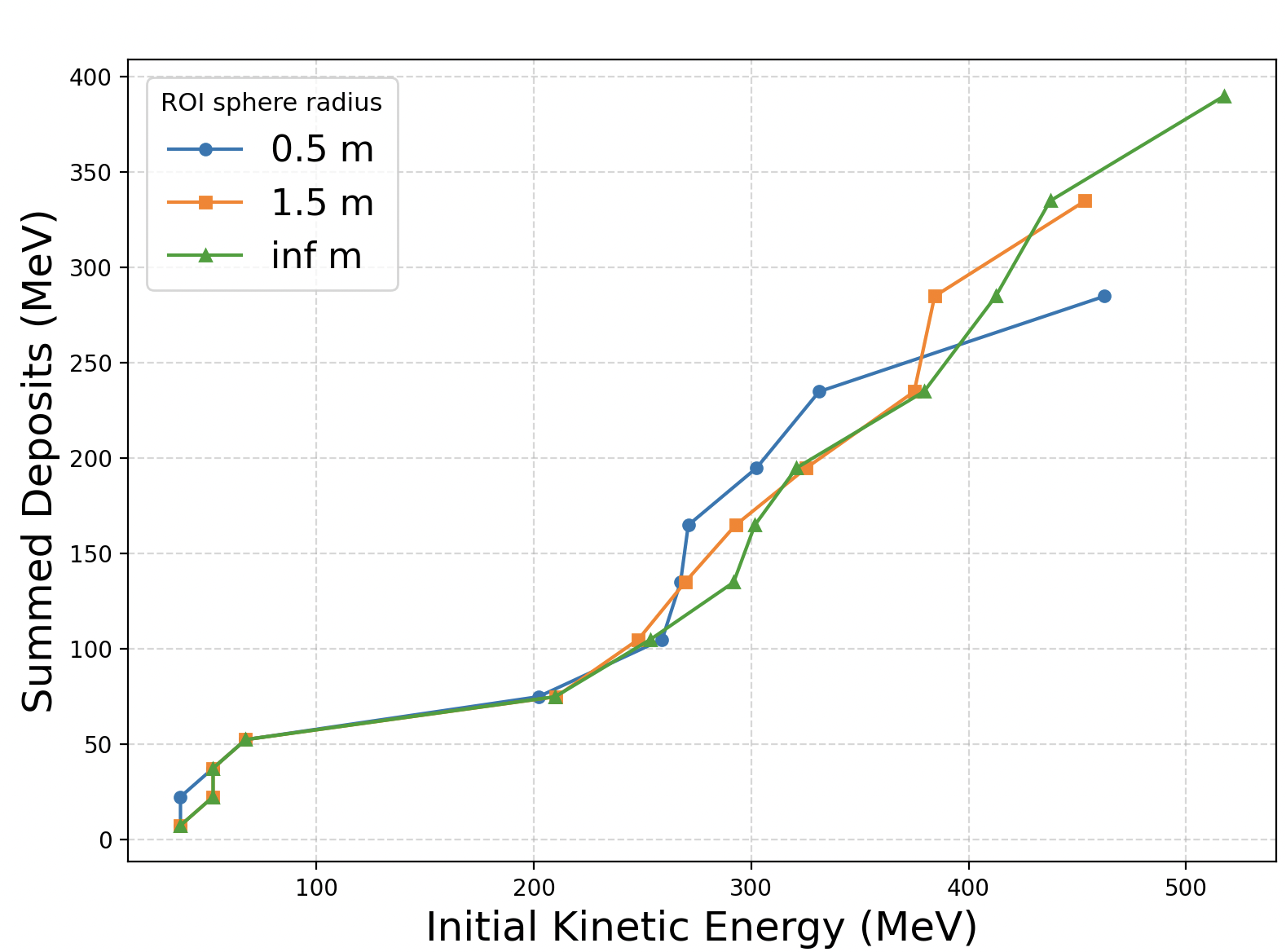}
    \caption{Neutron true kinetic energy relationship with summed energy deposit over different ROI volumes.}
    \label{fig:overlap}
\end{figure}

Naively, one would expect that with a larger ROI volume, more energy deposition will be included and it will more faithfully represent the true neutron kinetic energy. Here, we show that the radius plays a minor role in the energy analysis. In \cref{fig:neutronEcorrelation_radiusscan}, we show three scenarios of the spherical volume to sum up the neutron energy deposit. We consider the ideal case of infinite volume, a much smaller sphere with a 1.5 m radius, and another one with a 0.5 m radius. Despite significant energy smearing effects across sphere choices, the relationship between each slice of summed energy deposit and the corresponding most probable initial kinetic energy is similar. This relationship is indicated by the red dotted lines in \cref{fig:neutronEcorrelation_radiusscan}, and also separately plotted in \cref{fig:overlap}. This means that we can use a much smaller sphere to avoid including more noise and achieve a similar amplitude reconstruction performance. The impact of different spheres on the final reconstructed neutrino direction is also demonstrated in \cref{fig:money_plot_different_Esum_radius} in the next section.
 
\subsection{Neutrino direction reconstruction}
\label{sec:dir:nu-direction}

By integrating the neutron momentum direction and amplitude reconstruction detailed in previous sections, we demonstrate a substantial improvement in reconstructed neutrino directionality. To facilitate a comparative analysis, we define a baseline reconstruction, $R_{1}$, where all track-like particles are faithfully reconstructed. In this case, we use the true momenta of the final-state lepton and high-energy hadrons ($p$, $\pi^+$, $\pi^-$, and $\pi^0$). The kinetic energy threshold for reconstructing a proton remains 70 MeV, whereas for pions it is 20 MeV. This baseline ($R_{1}$) is then compared with neutron-included reconstruction, $R_{2}$, which augments the $R_{1}$ baseline with the reconstructed neutron momentum. Furthermore, we observe that the nuclear effects in the argon nucleus introduce an intrinsic angular deviation from the true neutrino direction, which means that there will be an angular bias even under perfect final-state particle reconstruction. We define this theoretically limited reconstruction $R_{3}$ as the vector sum of the true momenta of all final-state particles, representing the maximum achievable precision for neutrino directionality. 

\begin{figure}
    \centering
    \includegraphics[width=1.0\linewidth]{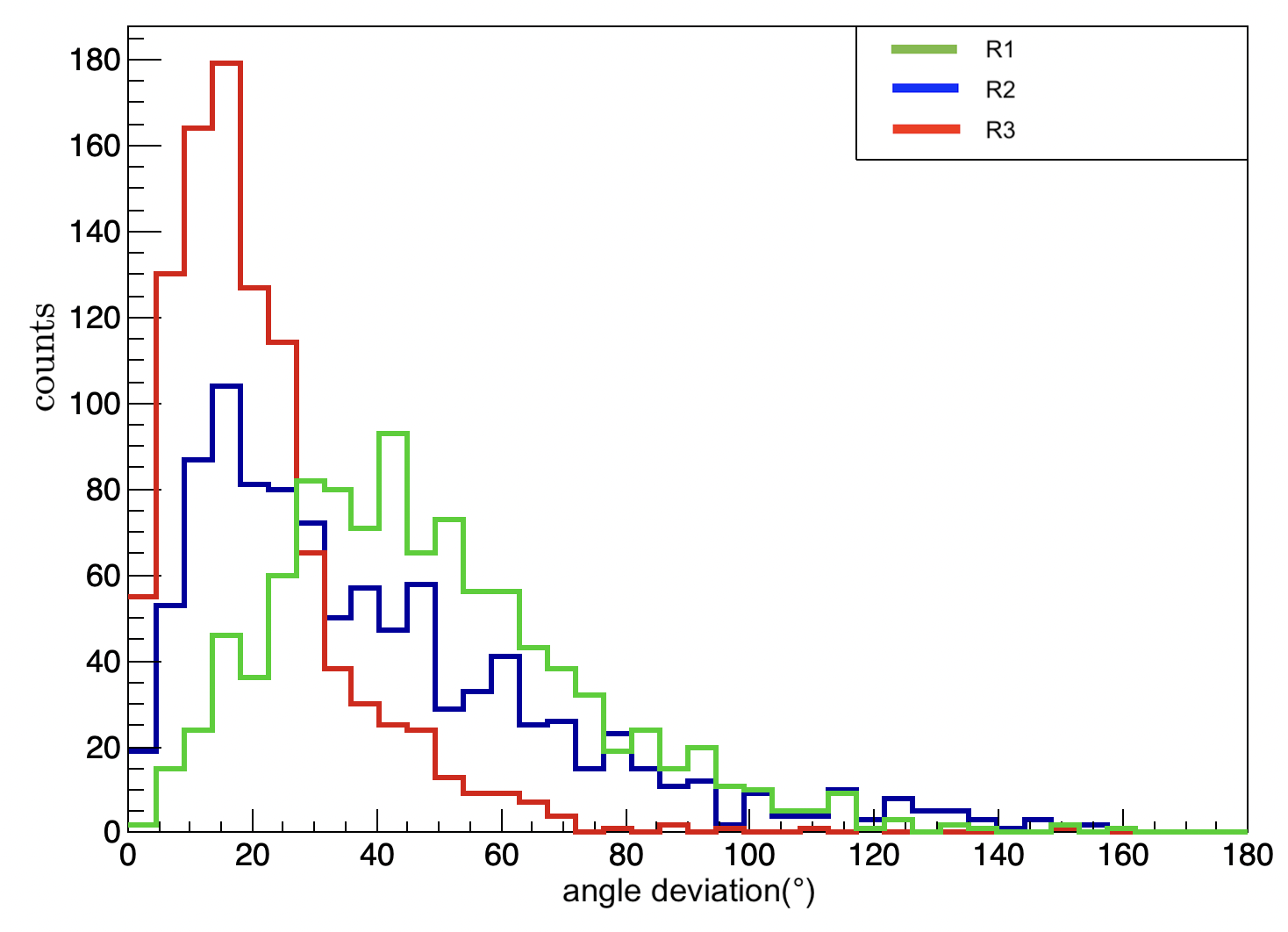}
    \caption{Angle difference between reconstructed neutrino direction and truth for the 0.5 GeV $\bar{\nu}_{\mu}$ sample. R1 (green) is the baseline reconstruction of all track-like particles. R2 (blue) is the reconstruction after adding the neutron reconstruction in this work to R1, and R3 (red) is the theoretically best reconstruction.}
    \label{fig:0p5GeVexampleplot}
\end{figure}

\begin{figure*}
    \centering
    \includegraphics[width=0.8\linewidth]{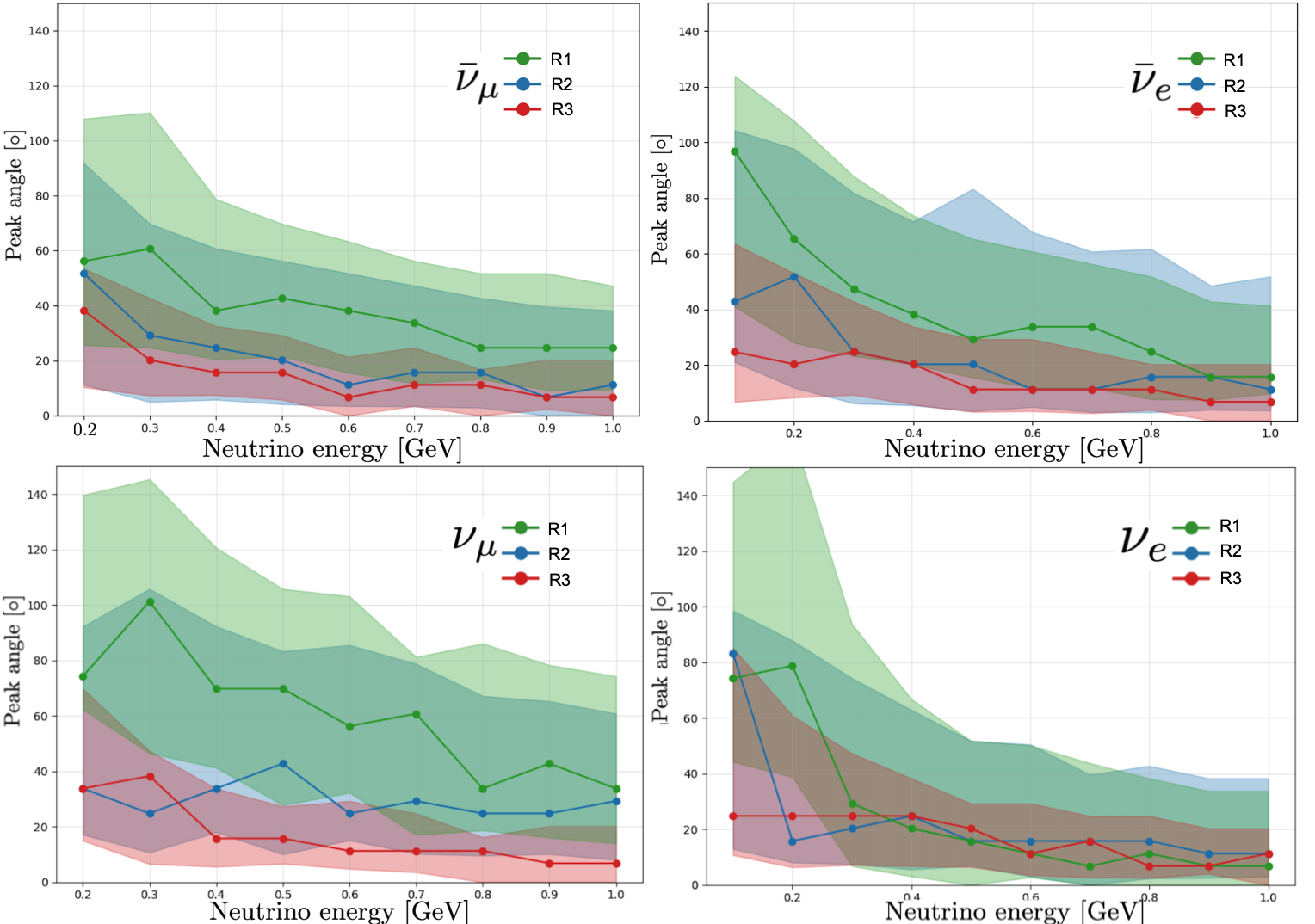}
    \caption{Neutrino direction reconstruction with a 1.5-meter radius sphere to capture the information for both direction and amplitude reconstruction.}
    \label{fig:money_plot_four_flavors}
\end{figure*}

Representative results for the neutrino direction reconstruction for a 0.5 GeV $\bar{\nu}_{\mu}$ sample are illustrated in \cref{fig:0p5GeVexampleplot}. The kinematic integration of the reconstructed neutron candidates ($R_{2}$) yields an angular-resolution distribution peak closer to 0$^\circ$, indicating enhanced directional reconstruction relative to the baseline ($R_{1}$). We characterize reconstruction performance using two primary metrics: the most probable angle and the 68\% percentile interval around it, which serves as the 1$\sigma$ statistical error bound. A comprehensive summary of the performance metrics, categorized by neutrino flavors and incident energies, is provided in \cref{fig:money_plot_four_flavors}. 

\begin{figure}
    \centering
    \includegraphics[width=\linewidth]{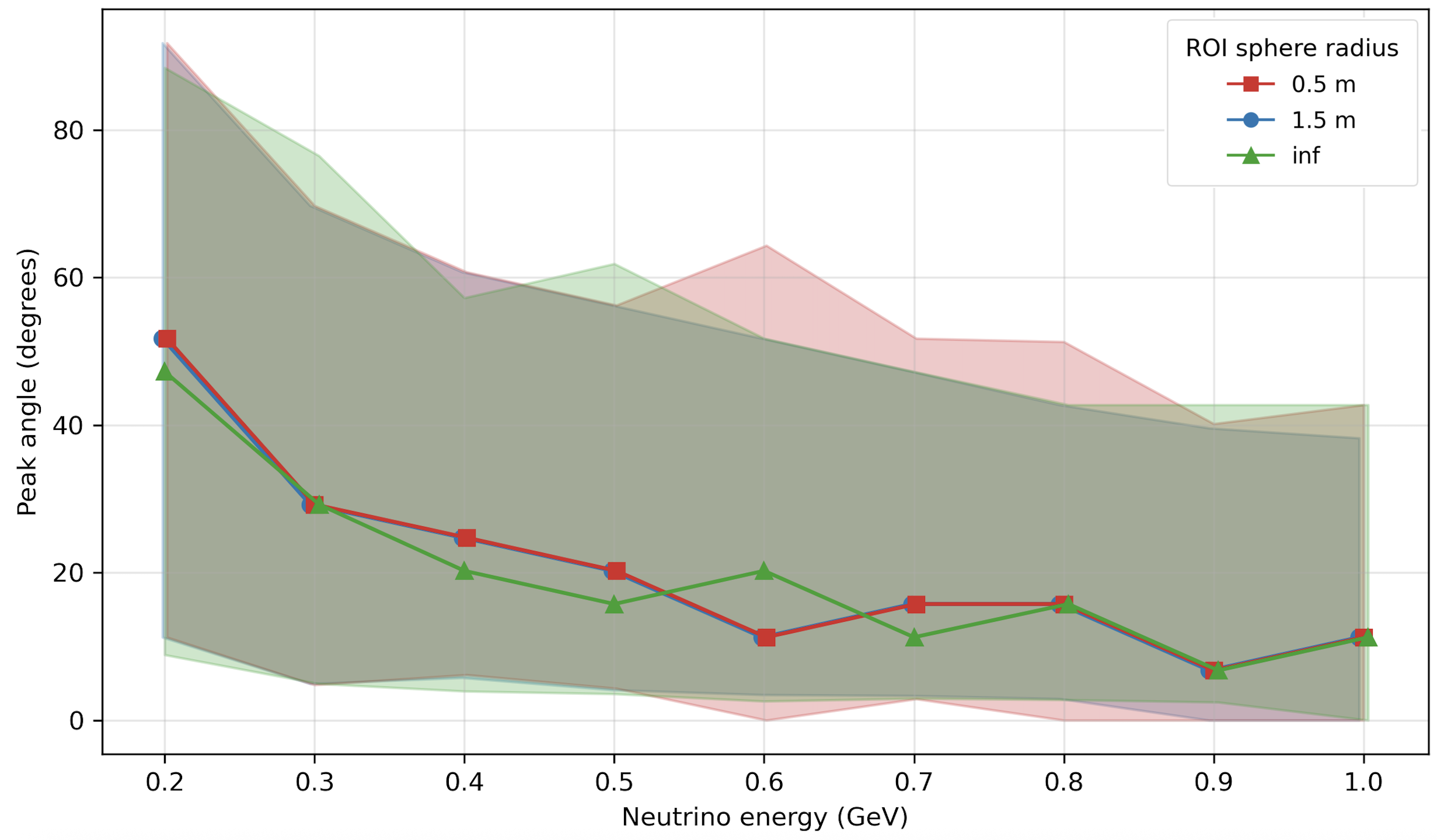}
    \caption{Performance of $\bar{\nu}_{\mu}$neutrino direction reconstruction of different ROI sphere radii: an infinite volume (green triangle), a sphere of 1.5 meters (blue circle), and one of 0.5 meters (red square).}
    \label{fig:money_plot_different_Esum_radius}
\end{figure}

In general, improvement in the most probable angular deviation is observed across neutrino flavors. The enhancement is most pronounced for antineutrinos, where the final-state neutron is kinematically required for complete neutrino momentum vector reconstruction. The fact that muons propagate as a track makes the separation between the muon and neutron energy deposition easier compared to the electron (anti)neutrino case. The integration of neutron reconstruction for electron (anti)neutrinos fails to yield a significant improvement in directional resolution, especially at higher neutrino energies. This lack of enhancement is primarily due to the kinematics of the high-energy regime, where the initial neutrino direction is strongly correlated with the outgoing electron (positron) shower, rendering additional neutron information redundant. Furthermore, the inherent stochastic uncertainties in neutron momentum reconstruction introduce noise that effectively impedes any potential gains in final directionality. We also found that varying the neutron-momentum-amplitude reconstruction setup from a 0.5-meter-radius sphere to an infinite volume does not significantly affect the ultimate neutrino reconstruction. This is demonstrated in \cref{fig:money_plot_different_Esum_radius}. Thus, it is safe to choose a smaller sphere for neutrino momentum reconstruction. This benefits a larger fiducial volume and minimal noise for future atmospheric neutrino analysis in LArTPCs.

\section{Discussion}
\label{sec:discuss}

The recent work in Ref.~\cite{n_blip_reco_Wan}, which details neutron blip reconstruction and its utility in sub-GeV neutrino physics, underscores the growing community interest and the importance of low-energy hadronic signatures in LArTPCs. A comparative review of these two works is imperative to provide a more holistic view of the physics limits and the path toward realistic detector implementation.

While aimed at a similar physics application, our neutrino direction reconstruction work is developed independently and is highly complementary to the study presented in Ref.~\cite{n_blip_reco_Wan}. Our study is designed as a first-principles physics analysis. By using energy deposits directly from \texttt{GEANT4} and applying fundamental detector threshold cuts to these deposits, without convolving any particle- or event-level energy resolutions, we establish the intrinsic physics limits on reconstruction performance. In contrast, Ref.~\cite{n_blip_reco_Wan} employs a more sophisticated simulation of the MicroBooNE detector response and evaluates performance utilizing a realistic atmospheric neutrino flux. While we acknowledge the modeling uncertainties inherent in neutrino interactions and fluxes, our work prioritizes the fundamental reconstruction potential of a LArTPC.

A key distinction lies in how the active volume is treated. While Ref.~\cite{n_blip_reco_Wan} utilizes a localized TPC volume, our study assumes an infinite LAr medium. This ensures that energy deposits are not truncated by detector boundaries, providing a complete accounting of the total neutrino energy profile deposited as charge and light signals, which is important for our study on energy reconstruction (\cref{sec:energy}) and charge separation (\cref{sec:separation}). Such an idealized containment study also serves as a useful benchmark for the design of future DUNE FD modules~\cite{DUNE_Phase2}.

The two works explore different strategies during reconstruction:
\begin{itemize}
  \item Directional Vectors: In our direction reconstruction (\cref{sec:dir}), we prioritize the most proximal energy deposit as the primary carrier of the neutron direction information, whereas Ref.~\cite{n_blip_reco_Wan} leverages the multiplicity of the blip system.
  \item Hadronic Treatment and Backgrounds: Ref.~\cite{n_blip_reco_Wan} performs a comprehensive analysis to discriminate neutron blips from other non-radioactive sources. Conversely, our framework treats the blip-like energy deposits collectively, incorporating contributions from all low-energy hadronic activity (including low-energy protons) without explicit discrimination. We adopt a more conservative $70\,\text{MeV}$ kinetic energy cut for proton detection to clearly delineate the boundary between track-like and blip-like objects. This approach allows the final performance metrics to naturally incorporate the effects of non-neutron hadronic noise.
  \item Thresholds: Both works adopt a $600\,\text{keV}$ threshold to simplify the treatment of the dominant $^{39}\text{Ar}$ radiological background. In \cref{sec:appA}, we demonstrate, using a first-principles analysis, that the $^{39}\text{Ar}$ background affects the final result only when the threshold is lowered to 300 keV or 100 keV. 
\end{itemize}

A major area of divergence and complementarity lies in the approach to neutrino charge separation. While Ref.~\cite{n_blip_reco_Wan} utilizes blip multiplicity to enhance reconstruction, our work focuses on the complementarity of charge and light signals. By using the combined energy deposits from both detector signals as inputs to a multivariate analysis, we achieve a $70\%$ separation efficiency for neutrino and antineutrino identification (\cref{sec:separation}). This demonstrates that even in the absence of detailed blip-type discrimination, the integrated calorimetric responses of both charge and light detectors provide a powerful tool for flavor and species tagging in the sub-GeV regime.

Some common findings emerge between the two studies regarding the challenges of $\nu_e$ reconstruction, in which the presence of electromagnetic showers complicates the isolation of neutron energy deposits (or blips). To address this, both frameworks employ similar geometric exclusion strategies (cones and cylindrical masks) to reject hadrons or de-excitation gammas below threshold near the vertex.

\section{Summary and Outlook}
\label{sec:summary}

Sub-GeV neutrinos are central to measurements of CP violation in the lepton sector, and the sensitivity of these measurements depends directly on how well their energy, direction, and charge can be reconstructed.

In this work, we studied \texttt{GEANT4} simulated energy deposits from sub-GeV neutrino charged-current interactions in an ideal LAr volume generated by \texttt{GENIE}. Many new insights have been gained into how its reconstruction could be improved in a LArTPC, which will inform atmospheric neutrino data analysis in upcoming LArTPC neutrino experiments such as DUNE.  

For the energy reconstruction, we established that the charge-plus-light (Q+L) energy reconstruction consistently offers the best resolution across the entire sub-GeV range. The light-only reconstruction can deliver a comparable resolution performance to Q+L. The light-only reconstruction is also significantly better than the charge-only reconstruction Q2, benefiting to some extent from the self-compensating light reported for multi-GeV neutrinos in Ref.~\cite{Selfcompensatinglight4GeV}. 

For neutrino charge-flavor separation, at least 70\% efficiency and purity are demonstrated using a simple multivariate analysis with inputs from calorimetric features of visible energy deposits in light and charge at different charge-detection thresholds.

We reconstructed the neutron direction by pointing from the neutrino vertex to the nearest isolated energy deposit, which is shown to be a good proxy for the first neutron interaction point. The neutron momentum magnitude is reconstructed by summing the isolated energy deposits around the neutrino vertex and is then projected to the true kinetic energy using a simulated 2D energy-relation map. Adding this reconstructed neutron momentum to all track-like particle momentum proves to effectively improve the peak angular deviation from 40 degrees to 20 degrees for the $\bar{\nu}_{\mu}$ sample. Less improvement is observed for ${\nu}_{e}$ or $\bar{\nu}_{e}$ samples because of the complications from background EM energy deposits.

Building on the direction reconstruction work, future work should focus on developing a realistic neutron reconstruction by incorporating a realistic detector simulation that includes charge-detection noise, physical energy-deposit backgrounds from radiologicals in bulk LAr, and detector materials. Advanced AI/ML tools can be explored to investigate possible improvements in reconstructing ${\nu}_{e}$ directions. For energy reconstruction and neutrino charge separation, future work should focus on calibrating the energy-scale response of both the charge and light signals. It is encouraging to see recent work on modeling light signals produced from neutron-argon interactions using a pulsed neutron source in small LArTPC prototypes~\cite{PNS_VDColdBox}. It makes a crucial step toward calibrating the energy scale for both charge and light using neutron captures on argon in bigger LArTPC volumes.

\bigskip

\begin{acknowledgments}

We thank L. Wan for the initial discussions that motivated this study. We also thank S. Dolan and S. Dytman for their insightful discussions on the event generator and tunes used in this paper. Through this project, S. Jain was first supported by NSF Research Experiences for Undergraduates and then the DUNE Training ExperienCe Hub (DUNE-TECH) programs.
\end{acknowledgments}

\appendix

\section{Separation in the Effect of the Charge Threshold}
\label{sec:appQ}

We can better understand the separation between the distributions of $Q_{500}/Q_{75}$ over $\nue$ and $\antinue$ events by expressing the overall $Q_{500}/Q_{75}$ ratio of an event as a weighted sum of the $Q_{500}/Q_{75}$ ratio for each category of primary particle in the event:
\begin{align*}
    \frac{Q_{500}}{Q_{75}}&= \frac{ Q_{500}^{(EM)} + Q_{500}^{(p)}+ Q_{500}^{(n)}+Q_{500}^{(\pi^\pm)}+...}{Q_{75}}
    \\&= 
\left(\frac{Q_{75}^{(EM)}}{Q_{75}}\right) \frac{Q_{500}^{(EM)}}{Q_{75}^{(EM)}}+ 
\left(\frac{Q_{75}^{(p)}}{Q_{75}} \right) \frac{Q_{500}^{(p)}}{Q_{75}^{(p)}}\\&+   
\left(\frac{Q_{75}^{(n)}}{Q_{75}} \right) \frac{Q_{500}^{(n)}}{Q_{75}^{(n)}}+   
\left(\frac{Q_{75}^{(\pi^\pm)}}{Q_{75}} \right) \frac{Q_{500}^{(\pi^\pm)}}{Q_{75}^{(\pi^\pm)}} +...
\end{align*}  
One expects that the distributions of the ratios $\frac{ Q_{500}^{(EM)} }{ Q_{75}^{(EM)} }$, $\frac{ Q_{500}^{(p)} }{ Q_{75}^{(p)} }$, $\frac{ Q_{500}^{(n)} }{ Q_{75}^{(n)} }$, etc. would be determined largely by properties of how EM-component particles ($e^\pm, \pi^0,\gamma$), protons, neutrons, etc. respectively deposit energy in LAr -- irregardless of what kind of interaction produced them -- and therefore be essentially the same between $\nue$ and $\antinue$ events in the same energy regime. 

\begin{figure}[h!]
    \centering
    \includegraphics[width=1\linewidth]{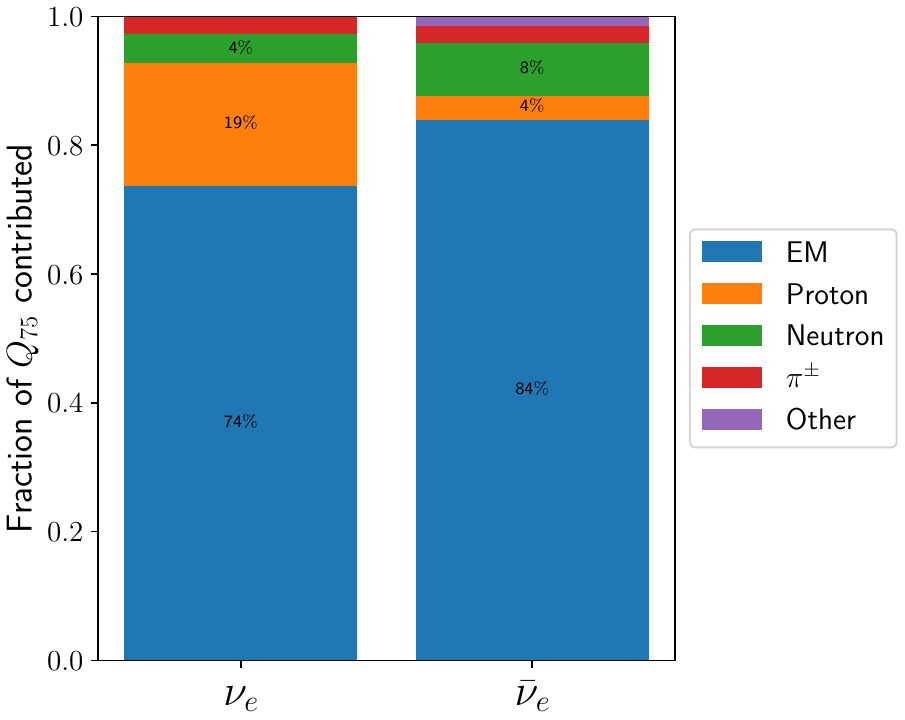}
    \caption{How the total energy-in-charge (as calculated with a 75 keV threshold) $Q_{75}$ is divided between different primary particle types on average across all $10^4$ simulated sub-GeV (discrete-$E_\nu$) $\nue$ and $\antinue$ events each. Note how the average share of $Q_{75}$ deposited by protons (and descendants) for $\antinue$ events is a fifth of what it is for $\nue$.}
    \label{fig:qdep_breakdown_flavor}
\end{figure}

The difference therefore must lie in the coefficients in the parentheses, which are the proportions of the event's total $Q_{75}$ apportioned to each category of primary particle. And, as expected from discussion around \cref{fig:edep_breakdown_flavor}, we find the largest discrepancy in the average $\frac{Q_{75}^{(p)}}{Q_{75}}$ i.e. the fraction of $Q_{75}$ resulting from primary protons: it is $19.1\%$ across the $\nue$ events but only $3.7\%$ across the $\antinue$ events as shown in \cref{fig:qdep_breakdown_flavor}. This is amplified by the fact that of the major primary particles, $Q$-deposits from protons are the least susceptible to the raising of the charge threshold (the average $\frac{ Q_{500}^{(p)} }{ Q_{75}^{(p)} }$ across all $\nue$ and $\antinue$ events is 0.94, whereas it is 0.48 and 0.28 for the EM-component and neutrons respectively). 

Indeed, if we exclude all depositions originating from primary protons, the $Q_{500}/Q_{75}$ distributions of the $\nue$ and $\antinue$ events actually become near-indistinguishable, suggesting that it is again the discrepancy in the prevalence of primary protons which the relationship between $Q_{500}$ and $Q_{75}$ exposes.

\section{$^{39}\text{Ar}$ decay background in bulk LAr}
\label{sec:appA}

$^{39}\text{Ar}$ decay in bulk LAr is a major noise source in neutron momentum reconstruction. They produce blip-like energy depositions, and it is difficult to distinguish them from neutron energy depositions. In the main text, we avoided this complication by raising the energy threshold to 0.6 MeV as mentioned in \cref{sec:dir:neutron-direction}. However, this threshold is high enough to eliminate some neutron energy-deposition information as well and potentially affect the reconstruction. 

Here, we discuss a more realistic case where isolated energy deposits, or blips, from both $^{39}\text{Ar}$ decay electrons and neutron interactions are registered in space and time close to the sub-GeV neutrino event in the LArTPC charge readout view, which is the starting point for many physics analyses. 

\begin{figure}
    \centering
    \includegraphics[width=1.0\linewidth]{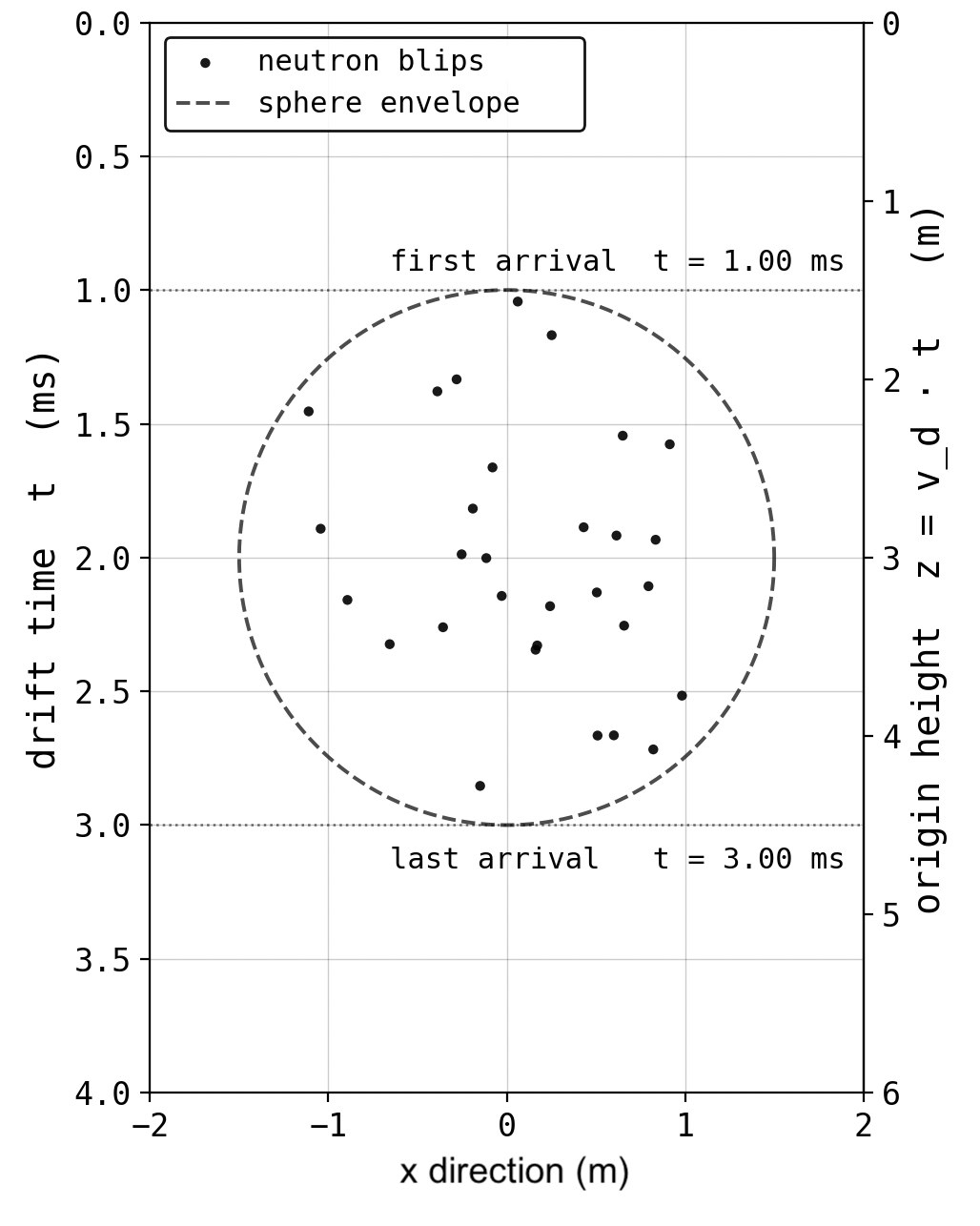}
    \caption{Illustration of an ROI volume of a neutrino event in the drift versus non-drift coordinates view. For a 1.5 m radius adopted in this analysis, the ROI size in the drift direction corresponds to about 2 ms in drift time, a characteristic time where $^{39}\text{Ar}$-induced noise blips from the full drift ($D$) should be considered. The total volume that we should consider for the $^{39}\text{Ar}$ noise is a cylinder volume with a radius of 1.5 m and a height of the full drift ($D$) of a TPC. This drawing has been produced with the assistance of Google Gemini.}
    \label{fig:chargereadoutview}
\end{figure}

We take a typical drift velocity $v_d$ $~\sim$ 1.5 $\text{ mm/}\mu\text{s}$ under a 500 V/cm electric field in LAr. As shown in the illustration in \cref{fig:chargereadoutview}, for a neutrino event occurring at $z_{vertex} = 3 \text{ m}$ with a ROI radius $R = 1.5 \text{ m}$ to look for isolated neutron energy deposits, this spatial ROI translates to a total charge readout time window $\Delta T_{total} = \frac{2R}{v_d} \sim$ 2 milliseconds in the charge readout view.

Now we estimate the $^{39}\text{Ar}$ noise contribution to this ROI. In the direction of drift, the $^{39}\text{Ar}$ in the bulk LAr along the full drift ($D$) of a LArTPC will contribute to the blips in the ROI over $\Delta T_{total}$. We assume that each electron from the decay of $^{39}\text{Ar}$ creates a single isolated blip. Using $^{39}\text{Ar}$ activity ($A$) of 1500 Bq/m$^{3}$, we have the total estimated number of blip noises from $^{39}\text{Ar}$ appearing in the ROI in the charge readout view:
\begin{equation}
\begin{aligned}
N_{\text{blips}} &= (\text{decay rate}) \times (\text{volume}) \times (\text{time}) \\
&= (\text{A}) \times (\pi R^2 \times D) \times \Delta T_{total} \\
&= \left( 1500 \, \frac{\text{decays}}{\text{s} \cdot \text{m}^3} \right) \times \left[ \pi (1.5 \, \text{m})^2 \times 6 \, \text{m} \right] \times (2 \times 10^{-3} \, \text{s}) \\
&\approx 127.
\end{aligned}
\end{equation}

\begin{figure}
    \centering
    \includegraphics[width=1.0\linewidth]{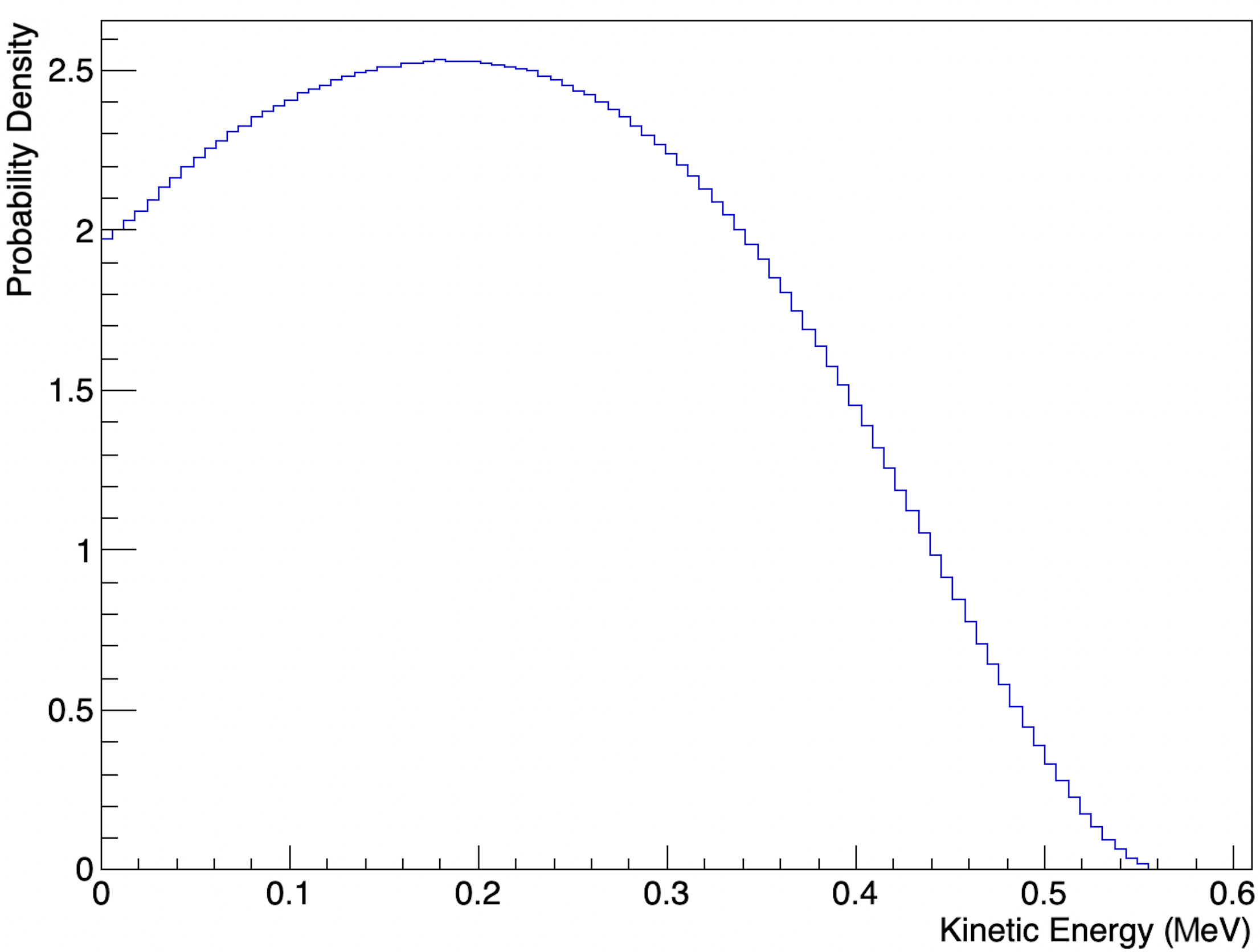}
    \caption{Normalized energy spectrum of $^{39}\text{Ar}$ decay electrons.}
    \label{fig:Ar39decayspectrum}
\end{figure}

Then we integrate over the $^{39}\text{Ar}$ decay energy spectrum shown in \cref{fig:Ar39decayspectrum}, and we find the number of blip-like noises with each energy threshold:
\begin{equation}
    \begin{aligned}
        N_{\text{blips}}(600\,\text{keV}) &= 0 \\
        N_{\text{blips}}(300\text{keV}) &\approx 37   \\
        N_{\text{blips}}      (100\,\text{keV}) &\approx 100
    \end{aligned}
\end{equation}
We randomly generate this number of points within the ROI defined above and repeat the study presented in the main text. 

\begin{figure}
    \centering
    \includegraphics[width=1.0\linewidth]{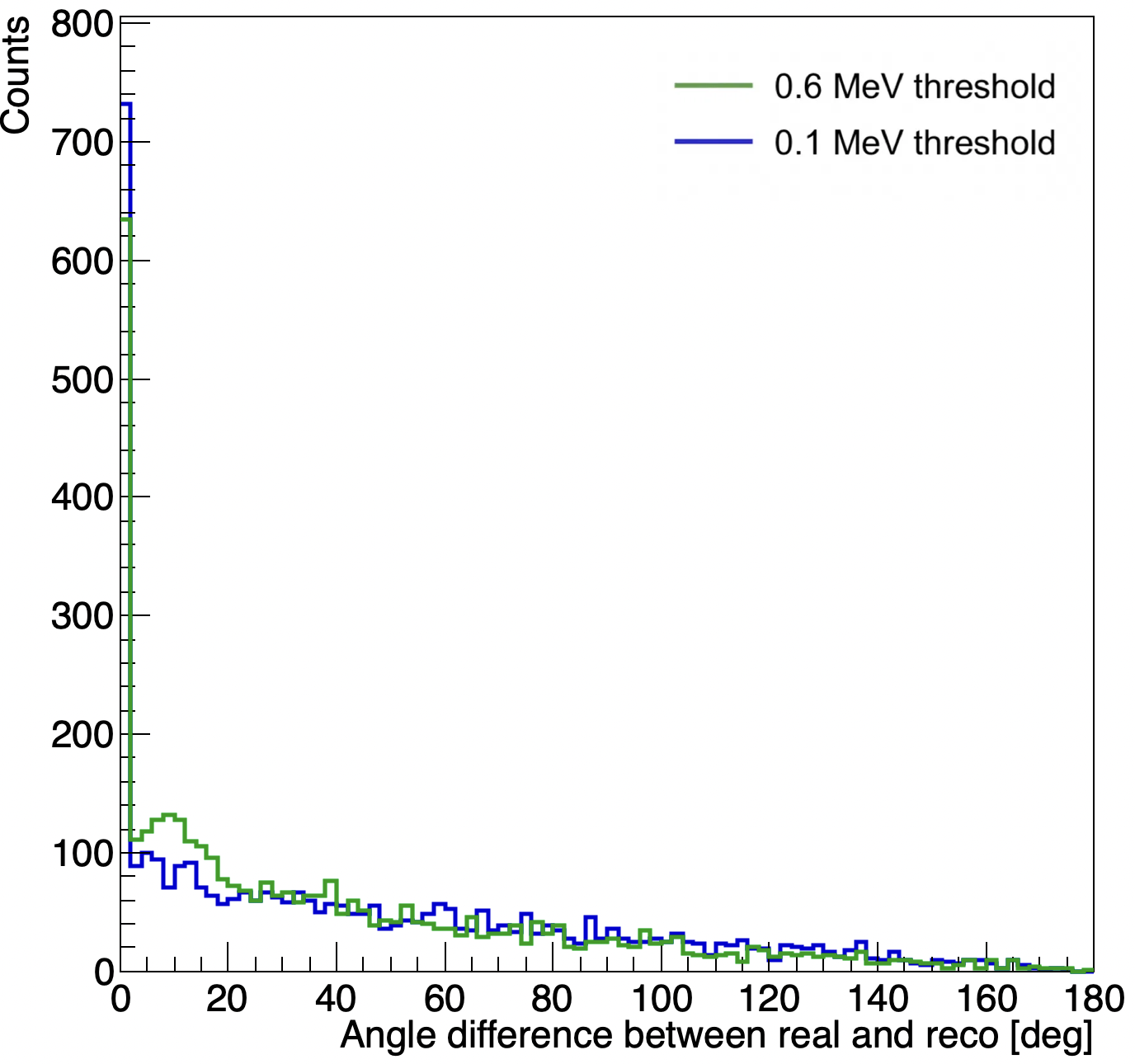}
    \caption{Performance of reconstruction of outgoing neutron direction with different energy thresholds applied to the same ROI volume (1.5 meter radius sphere from the neutrino vertex).}
    \label{fig:back_direction_reco}
\end{figure}

To understand the impact of background, we first examine neutron direction reconstruction. We consider the reconstruction of the events with one outgoing neutron and one outgoing muon in 0.5 GeV $\bar{\nu}_{\mu} $ events, and exclude all events with outgoing protons or other particles. This clean setup allows us to have a clear demonstration of the background influence on the neutron direction reconstruction. As the energy threshold for blip-like depositions is lowered, we capture lower-energy neutron depositions but also admit more background events. As shown in \cref{fig:back_direction_reco}, despite the overall performance being smeared by this increased background, the number of precisely reconstructed neutron directions improves. The near-perfect reconstruction indicated by the peak at 0° increases with decreasing threshold, driven by the emergence of lower-energy neutron-first interactions. However, in the 0–20° range, higher thresholds enable better direction reconstruction. While the reconstruction with a lower energy threshold exhibits a longer tail.

\begin{figure*}
    \centering
    \includegraphics[width=1.0\linewidth]{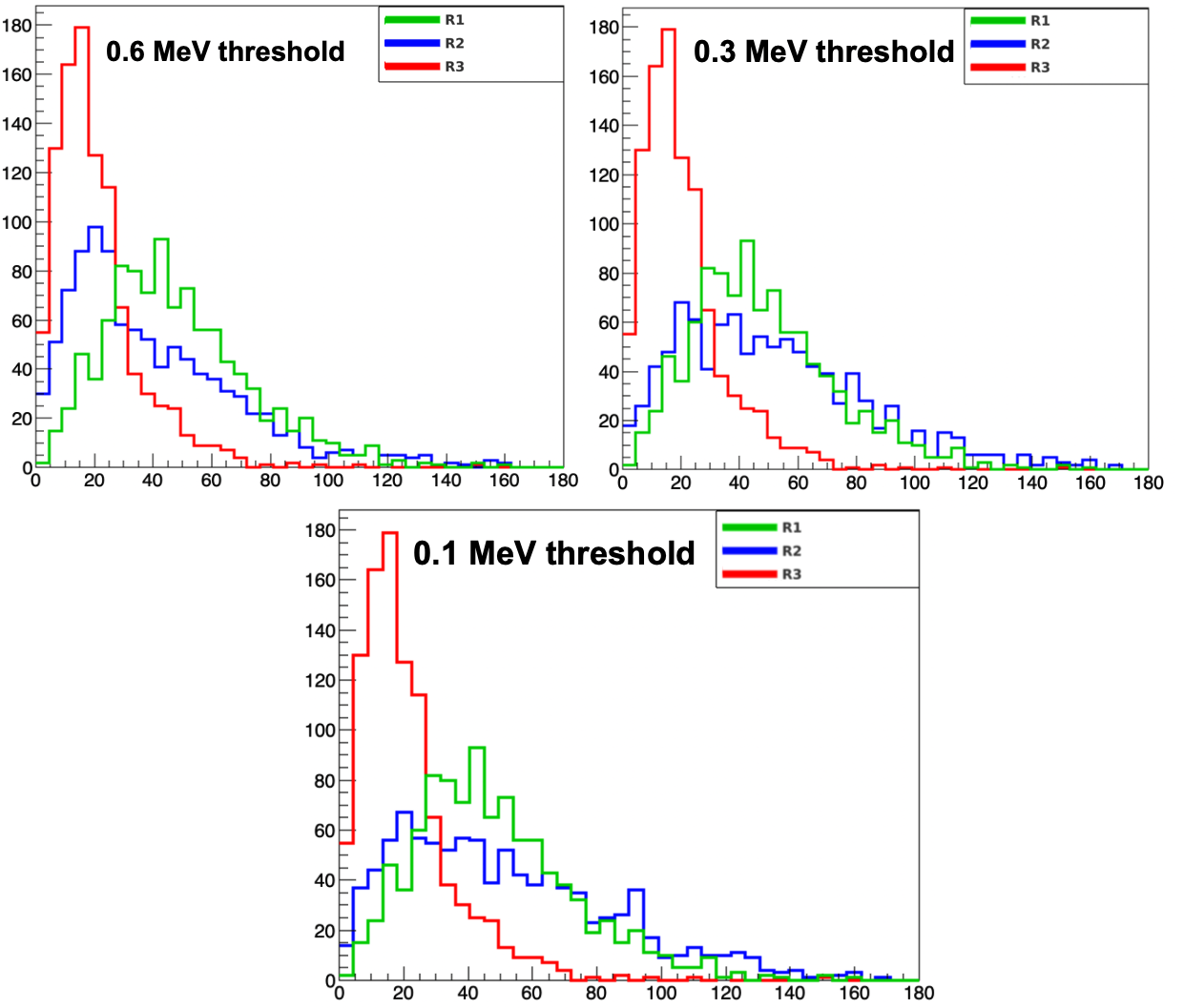}
    \caption{This shows the neutrino direction reconstruction performance obtained from 1000 events at $E_{\bar{\nu}_{\mu}}$=0.5 GeV. Three reconstruction schemes are compared: R1, the baseline reconstruction in which the final-state neutron momentum is neglected; R2, in which the reconstructed neutron momentum is included; and R3, the theoretical-best reconstruction obtained from truth-level kinematics in simulation. At a blip-like energy deposition threshold of 0.6 MeV, R2 yields a substantial improvement in angular resolution over R1, recovering an appreciable fraction of the gap between R1 and R3. As the threshold is lowered, however, this gain is progressively smeared out and becomes negligible, indicating that the benefit of including neutron-momentum information depends sensitively on the blip energy threshold.}
    \label{fig:moneyplot_slice}
\end{figure*}

\begin{figure}
    \centering
    \includegraphics[width=1.0\linewidth]{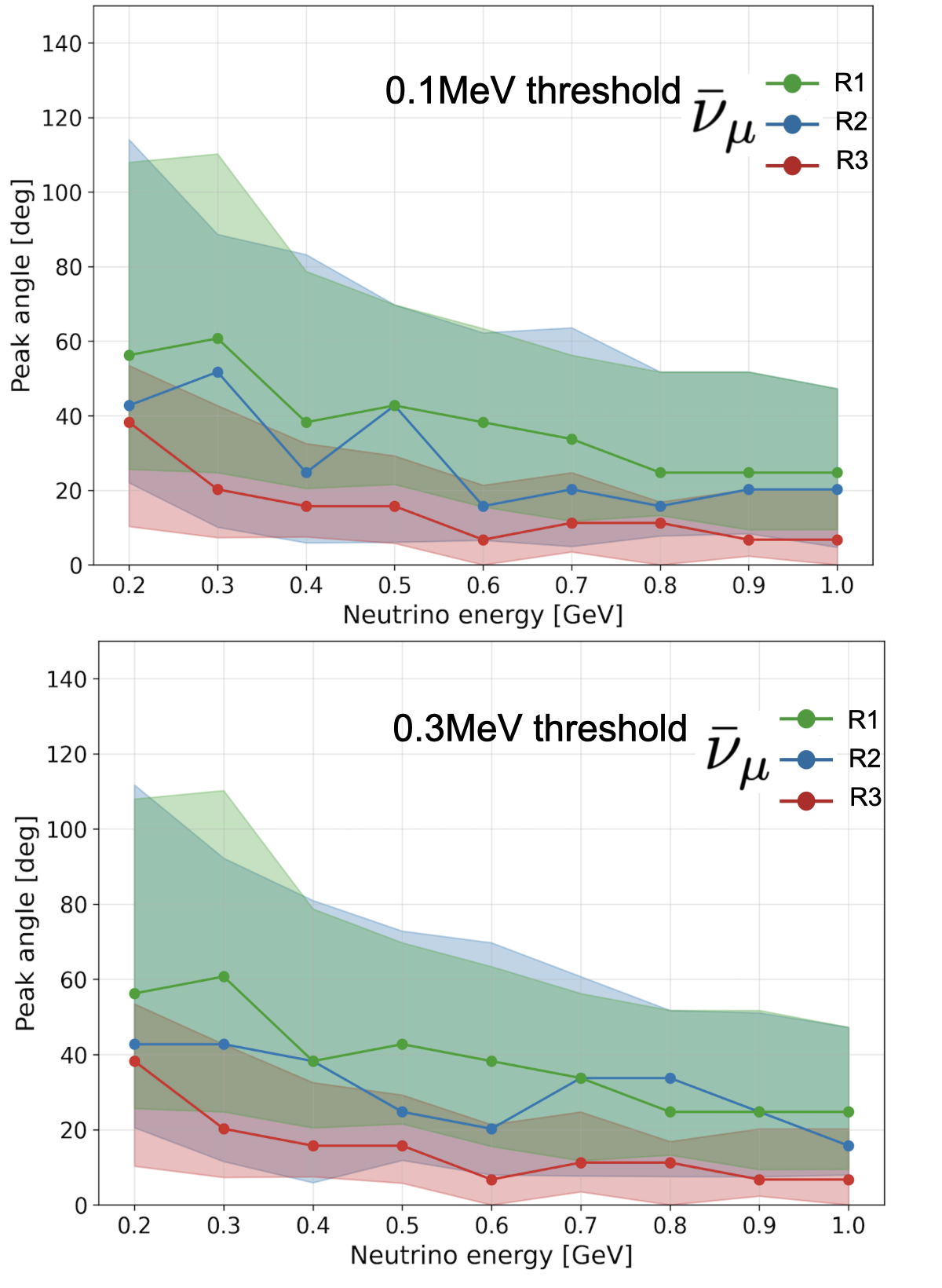}
    \caption{Neutrino direction reconstruction result with background included. We set the energy thresholds to 0.1 and 0.3 MeV, respectively, and assess the performance of R2, the reconstruction that uses the reconstructed neutron momentum, across different background levels.}
    \label{fig:moneyplot_withbackground}
\end{figure}

To compare the influence of the background and the low-energy threshold, we examine the final neutrino direction reconstruction result, shown in \cref{fig:moneyplot_slice}. Under the background assumptions discussed previously, lowering the energy threshold to 0.3 and 0.1 MeV significantly smears the original neutrino reconstruction distribution. The distribution that includes the neutron momentum reconstruction retains the same peak but develops a longer tail as the threshold decreases. This can be explained by the same effect in the neutron direction reconstruction shown in \cref{fig:back_direction_reco}.

The reconstruction results are presented in \cref{fig:moneyplot_withbackground} across neutrino energy bins in the $\bar{\nu}_{\mu}$ channel. The upper bound of the neutron-inclusive reconstruction is approximately consistent with the baseline, while the lower bound and the peak value remain below it. This behavior is in agreement with the 0.5~GeV reconstruction shown in \cref{fig:moneyplot_slice}, indicating a modest improvement in reconstruction performance even in the high-background regime.

\bibliography{main}
\end{document}